\pdfoutput=1

\pdfoutput=1
\documentclass[12pt]{article}
\usepackage[utf8]{inputenc}      % input font encoding
\usepackage{hyperref}
\usepackage{amsmath,amssymb,mathtools}
\usepackage[T1]{fontenc}          % output font encoding
\usepackage{booktabs,tabularx}
\usepackage{graphicx}
\usepackage{xspace}
\usepackage[usenames]{xcolor}\definecolor{fscolor}{RGB}{44,118,255}
\usepackage{geometry}
\usepackage{todonotes}
\usepackage{listings}
\usepackage[absolute]{textpos}
\usepackage[many]{tcolorbox}
\usepackage{xparse}
\usepackage[font=small,labelfont=bf,format=plain,margin=0.05\textwidth]{caption}
\usepackage{bbm}
\usepackage{tabularx}
\usepackage[normalem]{ulem}
\usepackage{soul}
\usepackage[capitalize]{cleveref}
\usepackage{slashed}
\usepackage{multirow}
\usepackage{subcaption}
\usepackage{pifont}
\usepackage{slashed}
\usepackage[shortlabels]{enumitem}
\usepackage[sorting=none,%
  citestyle=numeric-comp,%
  bibstyle=mynumeric,%
  giveninits=true]{biblatex}

\allowdisplaybreaks

\evensidemargin \oddsidemargin
\newcommand{\gev}{\;\text{GeV}\xspace}
\newcommand{\tev}{\;\text{TeV}\xspace}

\newcommand{\invfb}{\;\text{fb}^{-1}\xspace}

\newcommand{\Med}{{M}}
\newcommand{\Inv}{{I}}

\newcommand{\pTmiss}{\ensuremath{{p_\text{T}^\text{miss}}}\xspace}
\newcommand{\pTmissVec}{\ensuremath{p^{\text{ miss}}_\text{T}}\xspace}

\newcommand{\subfigsizetwo}{.3}

\NewBibliographyString{refname}
\NewBibliographyString{refsname}
\DefineBibliographyStrings{english}{%
  refname = {Ref\adddot},
  refsname = {Refs\adddot}
}
\DeclareCiteCommand{\ccite}
  {%
  \ifnum\thecitetotal=1
    \bibstring{refname}%
  \else%
    \bibstring{refsname}%
  \fi%
  \addnbspace\bibopenbracket%
  \usebibmacro{cite:init}%
   \usebibmacro{prenote}}
  {\usebibmacro{citeindex}%
   \usebibmacro{cite:comp}}
  {}
  {\usebibmacro{cite:dump}%
   \usebibmacro{postnote}%
   \bibclosebracket}
\newrobustcmd*{\Ccite}{\bibsentence\ccite}

\begin{document}

\thispagestyle{empty}
\def\thefootnote{\fnsymbol{footnote}}

\begin{flushright}
DESY-23-014\\
EFI-22-10
\end{flushright}
\vspace{3em}
\begin{center}
{\Large\bf A new LHC search for dark matter produced via\\[.2em] 
heavy Higgs bosons using simplified models}
\\
\vspace{3em}
{
Danyer Perez Adan$^a$\footnote{email: danyer.perez.adan@desy.de},
Henning Bahl$^b$\footnote{email: hbahl@uchicago.de},
Alexander Grohsjean$^c$\footnote{email: alexander.grohsjean@desy.de},
Victor Martin Lozano$^d$\footnote{email: victor.lozano@ific.uv.es},
Christian Schwanenberger$^{a,c}$\footnote{email: christian.schwanenberger@desy.de},
Georg Weiglein$^{a,c}$\footnote{email: georg.weiglein@desy.de},
}\\[2em]
{\sl ${}^a$ Deutsches Elektronen-Synchrotron DESY, Notkestr.\ 85, 22607 Hamburg, Germany}\\[1em]
{\sl ${}^b$ Department of Physics and Enrico Fermi Institute, University of Chicago,\\ 5720 South Ellis Avenue, Chicago, IL~60637~USA}\\[1em]
{\sl ${}^c$ Universit\"at  Hamburg, Luruper Chaussee 149, 22761 Hamburg, Germany\\[1em]
{\sl ${}^d$ Departament de Física Teòrica and IFIC, Universitat de València-CSIC,
E-46100, Burjassot, Spain}}

\def\thefootnote{\arabic{footnote}}
\setcounter{page}{0}
\setcounter{footnote}{0}
\end{center}
\vspace{2ex}
\begin{abstract}
{}

Searches for dark matter produced via scalar resonances in final states consisting of Standard Model (SM) particles and missing transverse momentum are of high relevance at the LHC. Motivated by dark-matter portal models, most existing searches are optimized for unbalanced decay topologies for which the missing momentum recoils against the visible SM particles. In this work, we show that existing searches are also sensitive to a wider class of models, which we characterize by a recently presented simplified model framework. We point out that searches for models with a balanced decay topology can be further improved with more dedicated analysis strategies. For this study, we investigate the feasibility of a new search for bottom-quark associated neutral Higgs production with a $b \bar b Z + \pTmiss$ final state and perform a detailed collider analysis. Our projected results in the different simplified model topologies investigated here can be easily reinterpreted in a wide range of models of physics beyond the SM, which we explicitly demonstrate for the example of the Two-Higgs-Doublet model with an additional pseudoscalar Higgs boson.

\end{abstract}

\newpage
\tableofcontents
\newpage
\def\thefootnote{\arabic{footnote}}

%%%%%%%%%%%%%%%%%%%%%%%%%%%%%%%%%%%%%%%%%%%%%%%%%%%%%%%%%%%%%%%%%%%%%%%%%%%%%%
%%%%%%%%%%%%%%%%%%%%%%%%%%%%%%%%%%%%%%%%%%%%%%%%%%%%%%%%%%%%%%%%%%%%%%%%%%%%%%

\section{Introduction}
\label{sec:intro}

The discovery of a scalar boson ten years ago marked a milestone for particle physics~\cite{ATLAS:2012yve,CMS:2012qbp}. Within the current experimental and theoretical uncertainties, it agrees remarkably well with the Standard Model (SM) Higgs boson. Besides further characterizing the discovered boson, searching for additional beyond the Standard Model (BSM) scalars is one of the main efforts within the LHC physics program.

Additional Higgs bosons could provide ways to address various unexplained experimental observations like dark matter (DM). Higgs bosons can constitute DM or act as a mediator between the visible and invisible sectors~\cite{Curtin:2013fra,Arcadi:2019lka}. Scenarios involving DM particles can be tested by direct DM detection experiments (searching for the scattering of DM particles with e.g.~nucleons), by indirect detection experiments (searching for the annihilation of DM particles), or collider experiments (searching for the production of DM particles at colliders)~\cite{Duerr:2016tmh,Feng:2010gw,Bertone:2004pz}.

Two complementary strategies are pursued at colliders to search for DM.
Besides the search for the decay of the mediator particle into DM particles, giving rise to a missing transverse momentum signature, searches are also performed for decays of the mediator particles into SM particles. Concerning the former type of searches, since the DM particles experimentally only manifest themselves in the form of a missing transverse momentum, the presence of at least one additional SM particle in the final state is required. 

Most of the ``mono-X plus missing momentum'' searches~\cite{ATLAS:2021kxv,CMS:2021far,ATLAS:2020uiq,CMS:2018ffd,ATLAS:2021gcn,CMS:2020ulv,ATLAS:2021shl,ATLAS:2021jbf,CMS:2019ykj,CMS:2018zjv,CMS:2018nlv,ATLAS:2022bzt,ATLAS:2020fgc,ATLAS:2022znu,ATLAS:2020yzc,CMS:2018ysw,CMS:2019zzl} --- with X being for example a jet, a photon, a Higgs boson, or a $Z$ boson 
--- are motivated by simplified scalar or vector portal models~\cite{Abdallah:2015ter,Abercrombie:2015wmb,LHCDarkMatterWorkingGroup:2018ufk}. In these portal models, a heavy resonance decays invisibly into dark matter particles and is produced in association with SM particles. 
As a result of this event topology, the missing transverse momentum recoils against the visible transverse momentum resulting in an ``unbalanced'' missing momentum distribution peaking at high values. 

In general, however, also other event topologies with a more ``balanced'' (i.e., softer) missing momentum distribution can appear in well-motivated BSM models. In order to classify the different decay topologies, a simplified model framework for scalar resonance searches with missing transverse momentum final states has been developed in \ccite{Bahl:2021xmy}, which allows one to cover a wider class of BSM models compared to a dedicated search within a particular model.

The aim of the present study is to demonstrate this approach by applying it to a potential search for DM in bottom-quark associated neutral Higgs production with a $b\bar b Z + \pTmiss$ (where \pTmiss is the missing transverse momentum) final state and perform a detailed collider study. This channel is so far not well explored and complementary to many existing $Z+\pTmiss$ final states searches, which often explicitly veto $b$-jets. Based on our analysis setup, we derive expected limits for the various simplified model topologies. In addition, as an application of our model-independent results we also provide expected limits for the Two-Higgs-Doublet model with an additional pseudoscalar DM portal (2HDMa).

This work is structured as follows. In \cref{sec:simp_model} we review the simplified model approach. The production of Monte-Carlo event samples and the event reconstruction is described in \cref{sec:mc_simulation}. We detail the event analysis in \cref{sec:analysis}. In \cref{sec:results}, we discuss the expected sensitivity of the proposed search. Our conclusions can be found in \cref{sec:conclusions}.

%%%%%%%%%%%%%%%%%%%%%%%%%%%%%%%%%%%%%%%%%%%%%%%%%%%%%%%%%%%%%%%%%%%%%%%%%%%%%%
%%%%%%%%%%%%%%%%%%%%%%%%%%%%%%%%%%%%%%%%%%%%%%%%%%%%%%%%%%%%%%%%%%%%%%%%%%%%%%

\section{Simplified models for mono-\texorpdfstring{$Z$}{Z} plus missing momentum final states}
\label{sec:simp_model}

The simplified model considered in the following represents a generic extension of the SM. Specifically, we focus on a scenario where a heavy scalar resonance decays into SM particles and transverse missing momentum with intermediate BSM states.\footnote{A detailed description of the simplified model approach for BSM Higgs searches can be found in \ccite{Bahl:2021xmy}.} The possible experimental signatures are characterised in terms of different simplified model topologies. The considered simplified model enlarges the SM with a heavy scalar resonance $\Phi$, a mediator $\Med$ and an invisible particle $\Inv$ (with masses $m_\Phi$, $m_\Med$, and $m_\Inv$, respectively). The spin nature of the mediator and the invisible particle could be either scalar, fermion or vector as described in \ccite{Bahl:2021xmy}. It is important to note that for the different topologies that can occur for the specified matter content we do not distinguish between the different types of mediators and between the different types of invisible particles. In this way the results obtained for the simplified model topologies can be mapped to different classes of explicit models. The approach of treating the mediator and the invisible particles in a generic way independently of their spin nature is motivated by the results in \ccite{Bahl:2021xmy} where it has been shown that the spin nature of the different particles has only a minor impact on the results for the different simplified model topologies.

\begin{figure}[tbh]\centering
%%%%%
\begin{subfigure}[t]{\subfigsizetwo\linewidth}\centering
\includegraphics[scale=1]{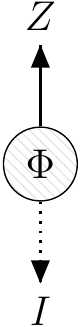}
\caption{1-vs-1 unbalanced}
\label{fig:monoZ_topologies_1vs1}
\end{subfigure}
%%%%%
\begin{subfigure}[t]{\subfigsizetwo\linewidth}\centering
\includegraphics[scale=1]{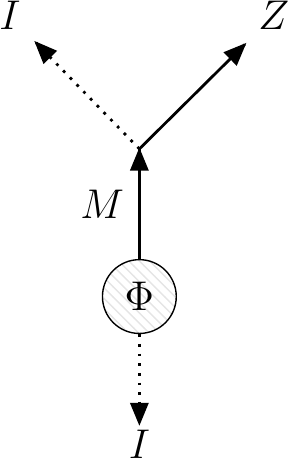}
\caption{2-vs-1 balanced}
\label{fig:monoZ_topologies_2vs1_balanced}
\end{subfigure}
%%%%%
\\[1em]
%%%%%
\begin{subfigure}[t]{\subfigsizetwo\linewidth}\centering
\includegraphics[scale=1]{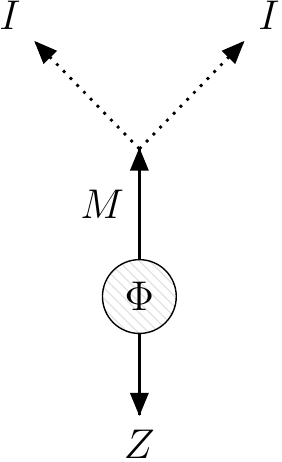}
\caption{2-vs-1 unbalanced}
\label{fig:monoZ_topologies_2vs1_unbalanced}
\end{subfigure}
\begin{subfigure}[t]{\subfigsizetwo\linewidth}\centering
\includegraphics[scale=1]{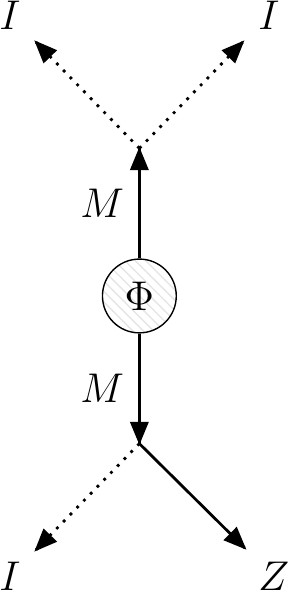}
\caption{2-vs-2 balanced}
\label{fig:monoZ_topologies_2vs2}
\end{subfigure}
\caption{Decay topologies of a neutral scalar boson $\Phi$ decaying in its rest frame to a $Z$ boson plus $\pTmiss$.}
\label{fig:monoZ_topologies}
\end{figure}

In this work, we will concentrate on the mono-$Z$ plus missing momentum signature (in association with additional $b$-jets from the production of the scalar resonance). A detailed study of this signature in the simplified model context can be found in \ccite{Bahl:2021xmy}. This final state gives rise to four different topologies that are shown in \cref{fig:monoZ_topologies} (where we omit the $b$-jets from the production of the scalar resonance). The first topology can be found in \cref{fig:monoZ_topologies_1vs1}. This topology is dubbed the 1-vs-1 unbalanced topology. In this case, the scalar resonance decays into a $Z$ boson and an invisible particle directly. As a consequence of the direct decay of the scalar, the invisible particle recoils against the $Z$ boson, resulting in a missing transverse momentum spectrum peaking at high $\pTmiss$ values. The second topology presented in \cref{fig:monoZ_topologies_2vs1_balanced} is the 2-vs-1 balanced topology. The decay products of the scalar resonance are a mediator and an invisible particle. The mediator subsequently decays into an invisible particle and a $Z$ boson. Given the fact that the scalar resonance is produced approximately at rest, the mediator will recoil against the invisible particle in the first step of the decay process, yielding a balanced missing transverse momentum spectrum. The third topology can be found in \cref{fig:monoZ_topologies_2vs1_unbalanced}. In this case, the scalar resonance decays into a $Z$ boson and a mediator, which decays into two invisible particles. As the $Z$ boson recoils against the mediator in the scalar reference frame, the $Z$ boson is produced in opposite direction with respect to the missing transverse momentum originating from the decay of the mediator. This results in a very similar kinematical situation as the 1-vs-1 topology, where the missing transverse momentum spectrum peaks at high $\pTmiss$ values. The last topology is the 2-vs-2 balanced topology, shown in \cref{fig:monoZ_topologies_2vs2}. In this case the scalar resonance decays into two mediator particles. One of them decays into a $Z$ boson and an invisible particle, while the second mediator decays into two invisible particles. Although it contains two mediators in the decay, this topology is similar to the 2-vs-1 balanced topology with respect to its kinematics. For that reason the missing transverse momentum spectrum features a similar balanced distribution in $\pTmiss$.

In \ccite{Bahl:2021xmy}, an additional topology for this specific signature has been discussed: the initial state radiation topology. In this case, the $Z$ boson is radiated from the initial state while the scalar resonance decays completely into invisible particles (directly or indirectly through different mediators). We do not consider this case in the present study because of the extremely low cross section resulting from requiring the presence of a $b\bar{b}$ pair accompanying the heavy resonance and the $Z$ boson as produced from an initial state radiation process. The details of the different Feynman diagrams contributing to each topology can be found in \ccite{Bahl:2021xmy}. 

In this work, we will concentrate on the four topologies shown in \cref{fig:monoZ_topologies} and perform a detailed collider analysis. Concretely, we focus on the production of a neutral scalar resonance via bottom-associated production and the subsequent decay to a $Z$ boson and invisible particles. No dedicated experimental search has so far been performed in this channel.

%%%%%%%%%%%%%%%%%%%%%%%%%%%%%%%%%%%%%%%%%%%%%%%%%%%%%%%%%%%%%%%%%%%%%%%%%%%%%%
%%%%%%%%%%%%%%%%%%%%%%%%%%%%%%%%%%%%%%%%%%%%%%%%%%%%%%%%%%%%%%%%%%%%%%%%%%%%%%

\section{MC samples and event reconstruction}
\label{sec:mc_simulation}

%%%%%%%%%%%%%%%%%%%%%%%%%%%%%%%%%%%%%%%%%%%%%%%%%%%%%%%%%%%%%%%%%%%%%%%%%%%%%%

This analysis is based on simulated $pp$ collisions at a center-of-mass energy of 13 TeV applying the conditions of the CMS detector~\cite{CMS:2008xjf} during the Run 2 data-taking period, in which an amount of data equivalent to a total integrated luminosity of 137 fb\textsuperscript{-1} was collected. Several SM background processes contribute to the signature explored in this work as detailed in \cref{sec:background-expectation} below. The most important background sources are $Z$+jets, $t\bar{t}$, single-top, and di-boson production in final states with two leptons. Monte Carlo simulated events are used to model the expected signal and background yields, as well as the relevant distributions of the observables that are used throughout the entire analysis.

Signal MC samples for the four topologies described in \cref{sec:simp_model} have been generated at leading order (LO) in perturbation theory using \textsc{MadGraph5\_aMC@NLO}~2.6.5~\cite{Alwall:2014hca,Frederix:2012ps} and the \textsc{UFO} model \texttt{simpBSM} provided in~\ccite{Bahl:2021xmy}. \textsc{Pythia}~8.230~ was used to simulate parton shower, hadronization, and the underlying event~\cite{Sjostrand:2014zea}. Signal samples have been generated varying two of the model mass parameters, resulting in a dedicated 2D scan per signal topology. All mass configurations have been chosen in such a way that none of the resonances ends up being off-shell in the decay chain. Therefore the mass scan is limited to the region of kinematically allowed $1 \rightarrow 2$ decays. More details regarding the mass scan implemented in each case are reported in Tab.~\ref{tab:signal-scan}. The selection of the mass parameter that is fixed in the scan is based on choosing the one whose variation does not change the kinematics or produces the 
smallest effect for the signature under 
consideration.\footnote{
For the case of the 1-vs-1 unbalanced topology, the mass of the mediator $m_{M}$ was formally set to $m_{M}=600$~GeV throughout this paper, however, this value has no impact on the results of this topology.}

\begin{table}[h]
  \begin{center}
    \renewcommand{\arraystretch}{1.}
    \begin{tabular}{|l||c|c|c|c|}
      \hline
      Topology & Mass fixed & Masses varied & Kin. constraints & No. points \\ 
      \hline\hline
      \multirow{1}{*}{ 1-vs-1 unbalanced } & \multirow{1}{*}{ $m_{M}$ } & \multirow{1}{*}{ $(m_{\Phi},m_{I})$ } & $m_{\Phi} \geq m_{I}+m_{Z} $ & \multirow{1}{*}{ 300 } \\
      \hline
      \multirow{2}{*}{ 2-vs-1 balanced } & \multirow{2}{*}{ $m_{\Phi}$ } & \multirow{2}{*}{ $(m_{M},m_{I})$ } & $m_{\Phi} \geq m_{I}+m_{M} $ & \multirow{2}{*}{ 80 } \\
      & & & $m_{M} \geq m_{I}+m_{Z} $ & \\
      \hline
      \multirow{2}{*}{ 2-vs-1 unbalanced } & \multirow{2}{*}{ $m_{I}$ } & \multirow{2}{*}{ $(m_{\Phi},m_{M})$ } & $m_{\Phi} \geq m_{Z}+m_{M} $ & \multirow{2}{*}{ 300 } \\
      & & & $m_{M} \geq 2m_{I} $ & \\
      \hline
      \multirow{3}{*}{ 2-vs-2 balanced } & \multirow{3}{*}{ $m_{\Phi}$ } & \multirow{3}{*}{ $(m_{M},m_{I})$ } & $m_{\Phi} \geq 2m_{M} $ & \multirow{3}{*}{ 28 } \\
      & & & $m_{M} \geq m_{I}+m_{Z} $ & \\
      & & & $m_{M} \geq 2m_{I} $ & \\
      \hline
    \end{tabular}
    \caption{The two-dimensional mass scans that have been performed for the four signal topologies. The second column indicates which mass parameter has been fixed in the scan, while the third one specifies the masses that were scanned for a given topology. The kinematic constraints taken into account when generating the individual mass grids are shown in the fourth column. In the last column the total number of mass points generated for each case is given.}
    \label{tab:signal-scan}
  \end{center}
\end{table}

The dominant background of $Z/\gamma^{*} \rightarrow ll$, denoted as Z+jets, has been generated at leading order with \textsc{MadGraph5\_aMC@NLO} interfaced to \textsc{Pythia} using the \textsc{MLM} matching scheme to properly describe hard emissions of up to four extra jets. The \textsc{Powheg}~v.2~\cite{Alioli:2010xd,Frixione:2007nw} generator interfaced to \textsc{Pythia} for showering is employed to generate top-quark pair production ($t\bar{t}$) and single-top processes at next-to-leading-order (NLO) in QCD. The various di-boson processes ($WW$, $WZ$, and $ZZ$) are produced at LO accuracy using \textsc{Pythia} as both matrix element and parton shower generator. 

The set of parton distribution functions (PDFs) used for simulating all the above samples is NNPDF 3.1 NNLO~\cite{NNPDF:2017mvq}, which have been accessed through the \textsc{LHAPDF} interface~\cite{Buckley:2014ana,Andersen:2014efa}. All background MC samples have been generated in the five flavour (\textsc{5F}) scheme, whereas the signal samples were produced in the \textsc{4F} scheme. The detector response simulation for all samples has been performed with the \textsc{Delphes}~3.5.0~\cite{deFavereau:2013fsa} package using the default configuration for the case of the CMS detector. For simplicity, simultaneous $pp$ collisions usually occurring in the same bunch crossing or in nearby bunch crossings, and commonly known as \textit{pileup}, are not considered in the simulation. The analysis described in the next section has been done within the framework of \textsc{MadAnalysis5}~1.8.45~\cite{Conte:2012fm}.

%%%%%%%%%%%%%%%%%%%%%%%%%%%%%%%%%%%%%%%%%%%%%%%%%%%%%%%%%%%%%%%%%%%%%%%%%%%%%%

%%%%%%%%%%%%%%%%%%%%%%%%%%%%%%%%%%%%%%%%%%%%%%%%%%%%%%%%%%%%%%%%%%%%%%%%%%%%%%
%%%%%%%%%%%%%%%%%%%%%%%%%%%%%%%%%%%%%%%%%%%%%%%%%%%%%%%%%%%%%%%%%%%%%%%%%%%%%%

\section{Analysis of the \texorpdfstring{$pp \to b\bar b \Phi (\to Z + \pTmiss)$}{pp -> bb Phi (-> Z + pTmiss)} process}
\label{sec:analysis}

%%%%%%%%%%%%%%%%%%%%%%%%%%%%%%%%%%%%%%%%%%%%%%%%%%%%%%%%%%%%%%%%%%%%%%%%%%%%%%

\subsection{Event selection}
\label{sec:event-selection}

The event selection targets a signal topology in which a substantially boosted $Z$ boson is produced in association with a pair of relatively forward $b$-jets. Due to the invisible particles in the final state, events are moreover characterized by large missing transverse momentum, \pTmiss. The selection of the final state particles, i.e.\ the leptons ($e$ and $\mu$), jets and \pTmiss, is designed to be as inclusive as possible with respect to the scanned mass points.

Electrons and muons are selected if they fall within the pseudorapity range $|\eta| < 2.4$, and if they both have at least a transverse momentum of $p_{\mathrm T}>10$ GeV, here called \textit{loose leptons}. Signal leptons, referred to as \textit{leptons}, are expected to be more energetic due to the boosted $Z$ boson, hence a higher cut of $p_{\mathrm T}>20$ GeV is applied. Additionally, an isolation requirement is added for both electrons and muons. For electrons the isolation is calculated using the energy deposited in a cone of $\Delta R < 0.3$ around the electron and required to be less than 15\% of the electron energy. For muons the isolation is calculated from the energy of all tracks in a cone of $\Delta R < 0.3$ around the muon and required to be less than 10\% of the muon energy.

Selected jets must have a minimum transverse momentum of $p_{\mathrm T}>20$ GeV. Two main jet definitions are used in the analysis; the \textit{standard} jets, which are additionally required to have a pseudorapidity value satisfying $|\eta| < 2.4$, and the \textit{forward} jets, allowed to reach pseudorapidity values of up to $|\eta| < 5$. The latter are relevant given the marked presence of jets with high pseudorapidity in the signal processes. The standard jets are further labeled depending on whether they pass the $b$-tagging algorithm or not. If they pass the criterion, they are called \textit{b-tagged} jets. All jets are cleaned by requiring the absence of any type of loose lepton inside a cone of $\Delta R < 0.4$ around the jet momentum.

The base event selection starts by requiring the presence of exactly two leptons of same flavour  and vetoing events with additional loose leptons. Selected leptons must be oppositely-charged, and their invariant mass must be within a window of $76\,\text{GeV} < m(l^{+}l^{-}) < 106\,\text{GeV}$, which corresponds to the bulk of the distribution of the reconstructed $Z$ boson mass. In order to reduce the $VV$ and $Z$+jets backgrounds while keeping almost the entire signal, a cut on the transverse momentum of the leading lepton of $p_{\mathrm T}(l_{\text{lead}})>50$ GeV is applied, as well as on the transverse momentum of the di-lepton system with the same lower threshold of $p_{\mathrm T}(l^{+}l^{-})>50$ GeV. Moreover, the angular separation between the two leptons is required to be $\Delta R(l^{+},l^{-}) < 3$. Given that in most signal scenarios the largest amount of \pTmiss is usually produced when the invisible particles are back-to-back with respect to the $Z$ boson, a moderate lower threshold for the difference in azimuthal angle between \pTmissVec and the di-lepton system is included in the selection, applying $\Delta \phi(\vec{p}^{\text{ miss}}_{\mathrm T},l^{+}l^{-}) > 0.5$. In all signal topologies there is a relatively heavy intermediate resonance ($\phi$). Thus, a large reconstructed invariant mass of its decay products is expected. However,  as it is only possible to estimate the missing momentum in the transverse plane, the transverse mass of the \pTmissVec and di-lepton systems, partly encoding the information about the mass of $\phi$, is used instead in order to further reduce the SM background, imposing $m_{T}(\vec{p}^{\text{ miss}}_\text{T},l^{+}l^{-}) > 140\,\text{GeV}$. 

Two signal regions are constructed based on the information coming from the jets in the event. The first region, named as \textit{Standard-SR}, is defined by requiring at least one $b$-tagged jet in the event on top of the above base selection. The other signal region, named \textit{ForwardJets-SR}, is constructed by demanding no $b$-tagged jets but at least two forward jets with a separation in pseudorapidity of $|\eta(j_{1})-\eta(j_{2})|>2.5$. If more than two forward jets are found in the event, the pair combination with the largest value for this quantity is considered for the above condition. No requirement is imposed on the \pTmiss, given that the full distribution is employed in the statistical analysis, as it will be explained in detail later. A summary of all the kinematic selections described above can be found in Table~\ref{tab:selection-summary}.

\begin{table}[h]
  \begin{center}
    \renewcommand{\arraystretch}{1.4}
    \begin{tabular}{|l||cc|}
      \hline
      Quantity & Standard-SR & ForwardJets-SR \\ 
      \hline\hline
      $N_{l}$ (opposite-charge, same-flavour) & \multicolumn{2}{c|}{$=2$ (with additional lepton veto)} \\
      $p_\text{T}(l)$ & \multicolumn{2}{c|}{$50/20$ GeV leading/trailing} \\
      $m(l^{+}l^{-})$ & \multicolumn{2}{c|}{$76\,\text{GeV} < m(l^{+}l^{-})  < 106\,\text{GeV}$} \\
      $p_\text{T}(l^{+}l^{-})$ & \multicolumn{2}{c|}{$>50$ GeV} \\
      $\Delta R(l^{+},l^{-})$ & \multicolumn{2}{c|}{$<3$} \\
      $\Delta \phi(\vec{p}^{\text{ miss}}_\text{T},l^{+}l^{-})$ & \multicolumn{2}{c|}{$>0.5$} \\
      $m_\text{T}(\vec{p}^{\text{ miss}}_\text{T},l^{+}l^{-})$ & \multicolumn{2}{c|}{$>140$ GeV} \\
      $N_{\text{b-tag}}$ & $\geq 1$& $=0$ \\
      $|\eta(j_{1})-\eta(j_{2})|_{\text{max}}$ & $-$& $>2.5$ \\
      \hline
    \end{tabular}
    \caption{Summary of the kinematic selections for the two defined signal regions.}
    \label{tab:selection-summary}
  \end{center}
\end{table}

%%%%%%%%%%%%%%%%%%%%%%%%%%%%%%%%%%%%%%%%%%%%%%%%%%%%%%%%%%%%%%%%%%%%%%%%%%%%%%

\subsection{Signal selection efficiency}
\label{sec:efficiency-maps}

\begin{figure}
    \centering
    \includegraphics[width=0.49\textwidth]{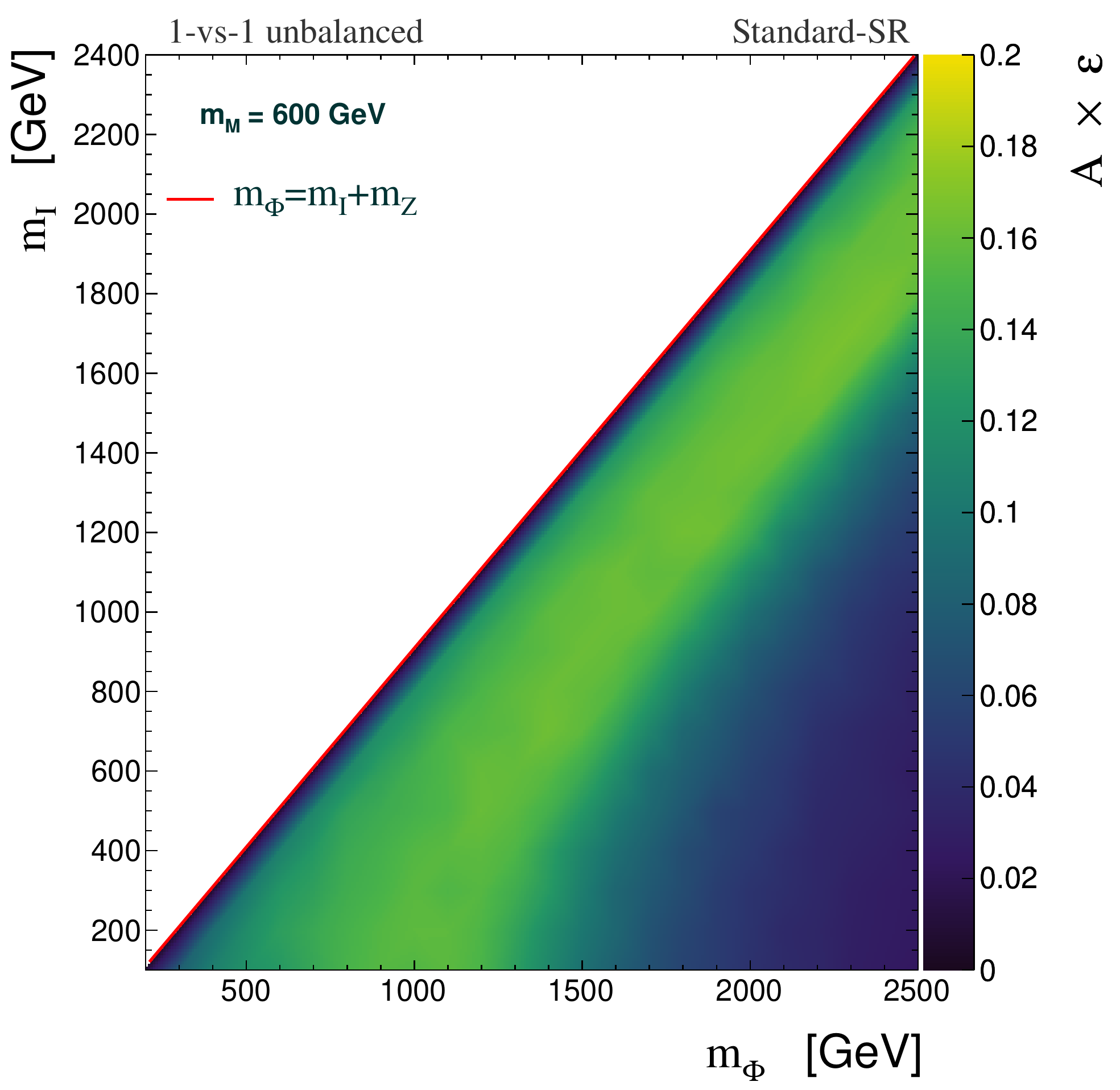}
    \hfill
    \includegraphics[width=0.49\textwidth]{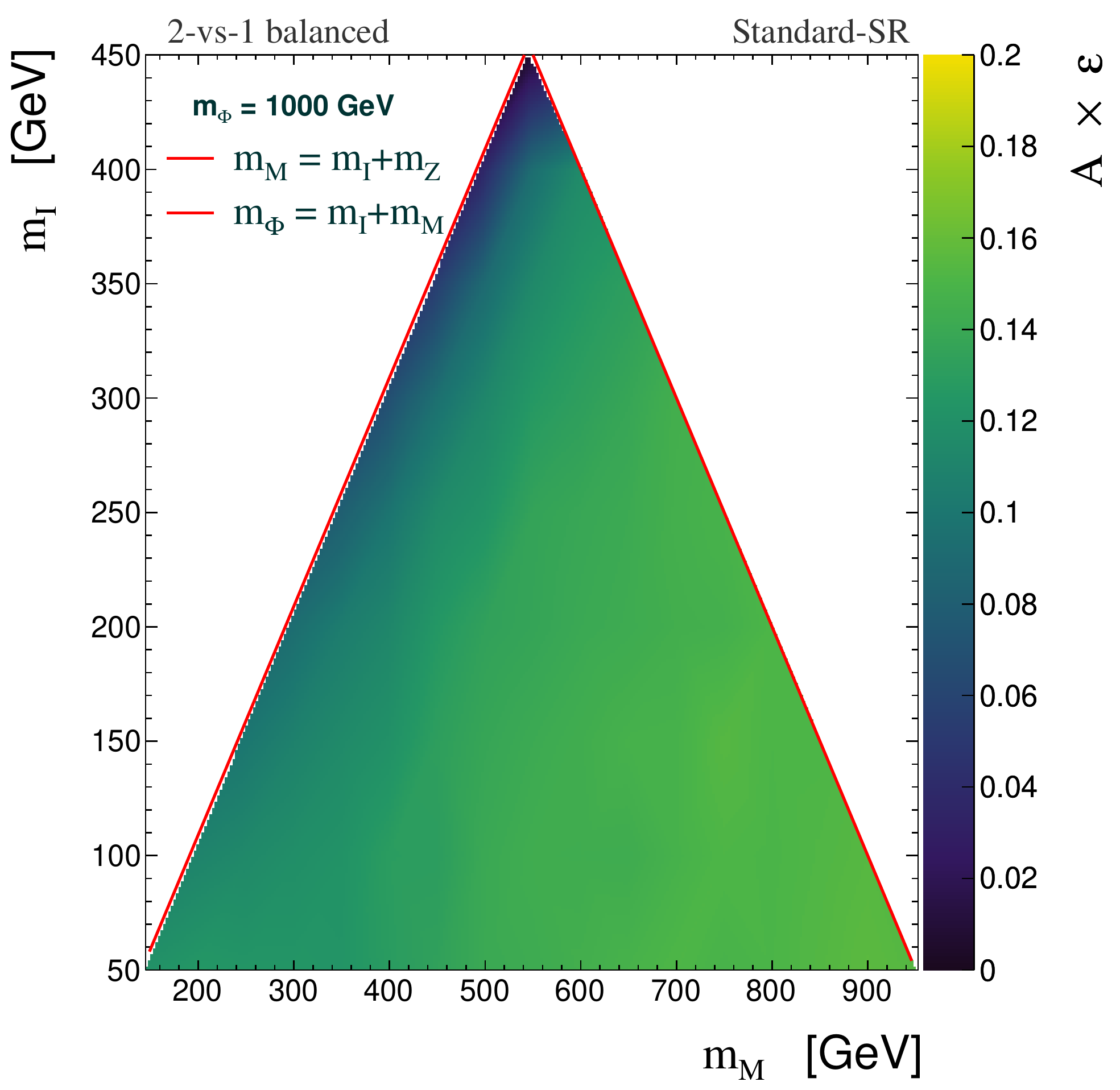} \\
    \vspace{1cm}
    \includegraphics[width=0.49\textwidth]{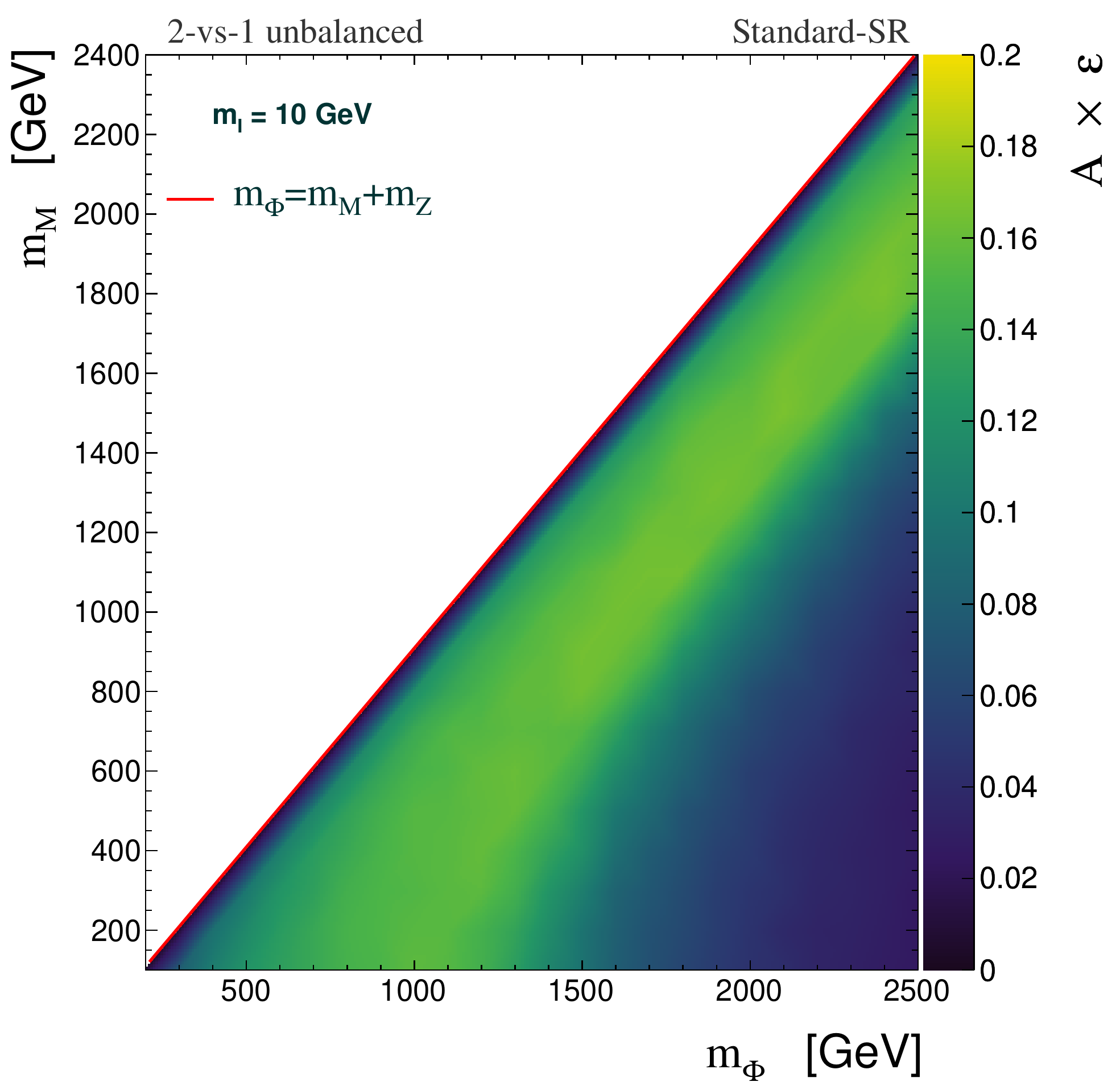}
    \hfill
    \includegraphics[width=0.49\textwidth]{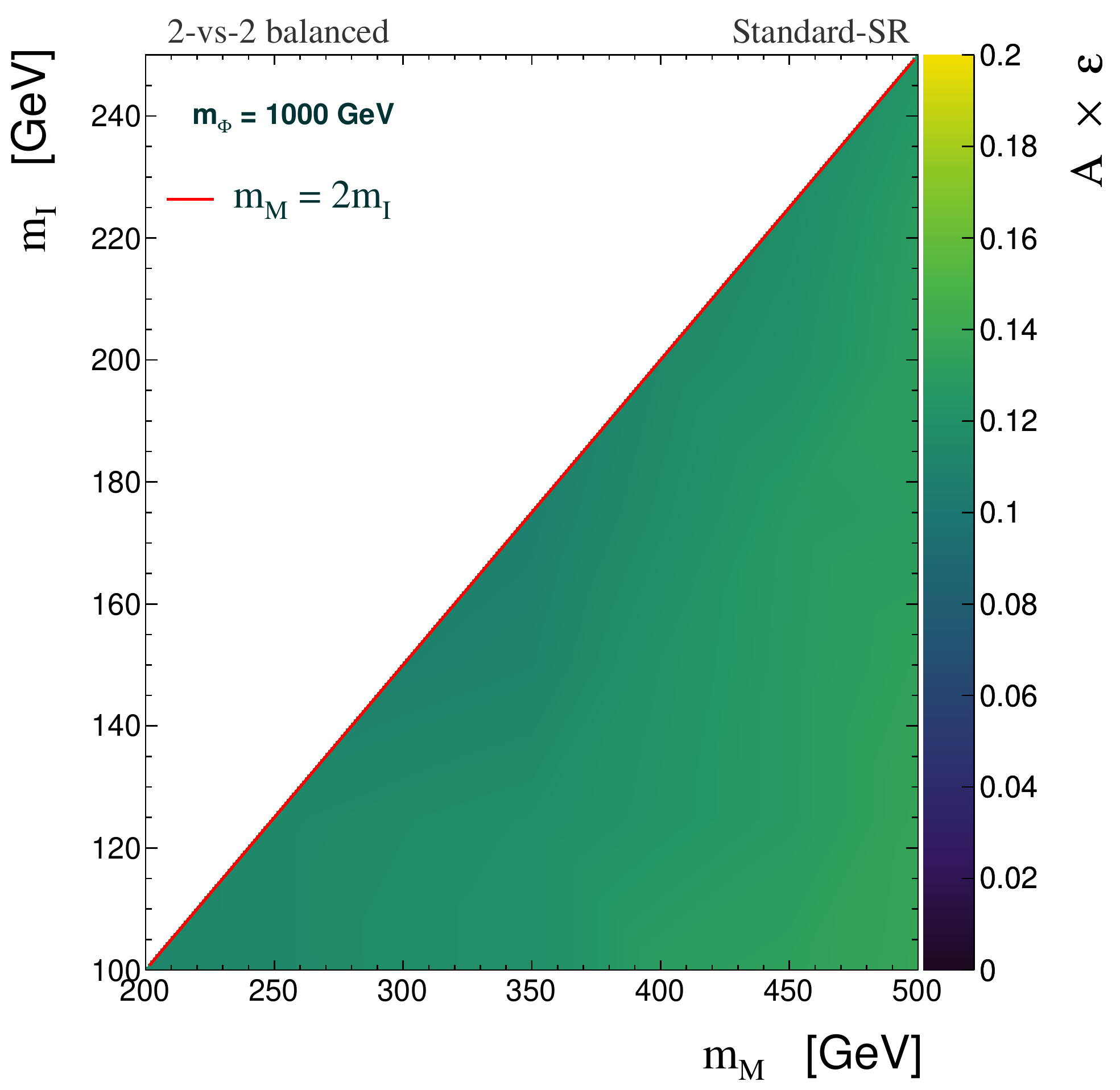}
    \caption{Analysis efficiency (including detector acceptance) for the selection corresponding to the Standard-SR (for a sample produced with the charged lepton decays of the Z boson). 
    The horizontal and vertical axes indicate the mass combination for which the efficiency is measured, while the color coding indicates the efficiency value obtained for this mass combination. The results are shown for the 1-vs-1 unbalanced (upper left panel), 2-vs-1 balanced (upper right panel), 2-vs-1 unbalanced (lower left panel), and 2-vs-2 balanced (lower right panel) topologies. The kinematic constraints for each topology are indicated by solid red lines, whereas the fixed mass value is specified in the legend.}
    \label{fig:Efficiency-Standard-SR}
\end{figure}

\begin{figure}
    \centering
    \includegraphics[width=0.49\textwidth]{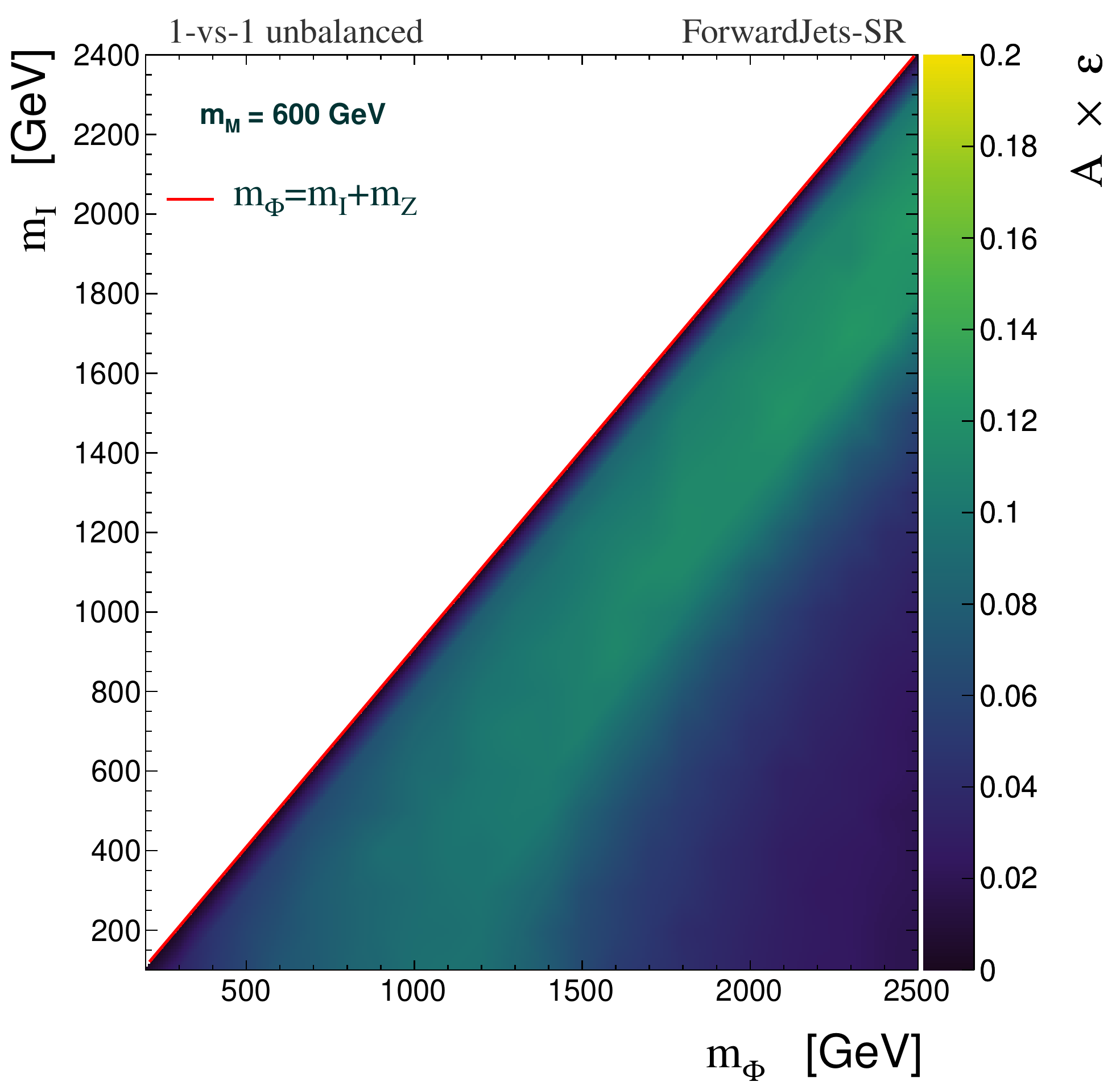}
    \hfill
    \includegraphics[width=0.49\textwidth]{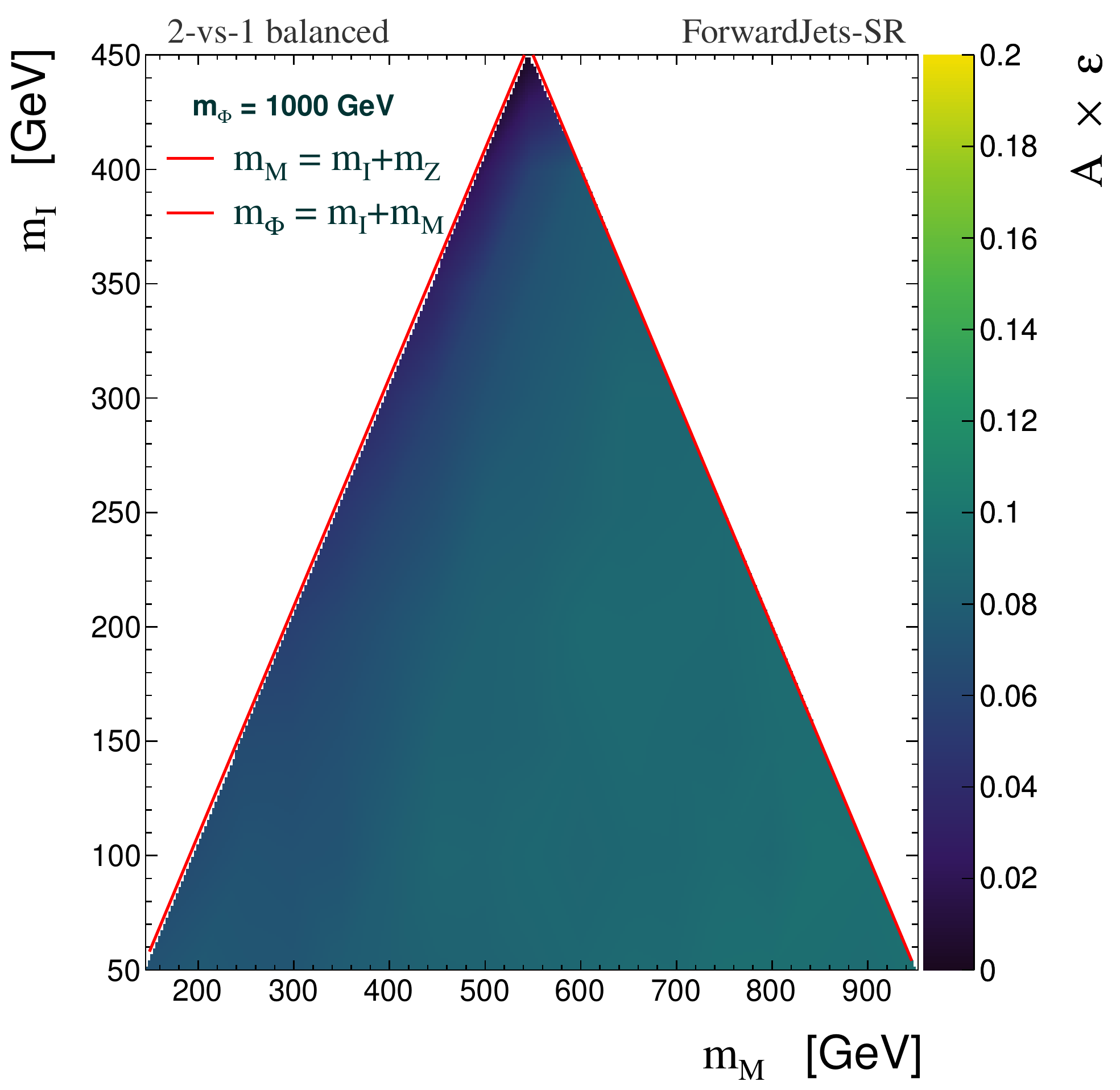} \\
    \vspace{1cm}
    \includegraphics[width=0.49\textwidth]{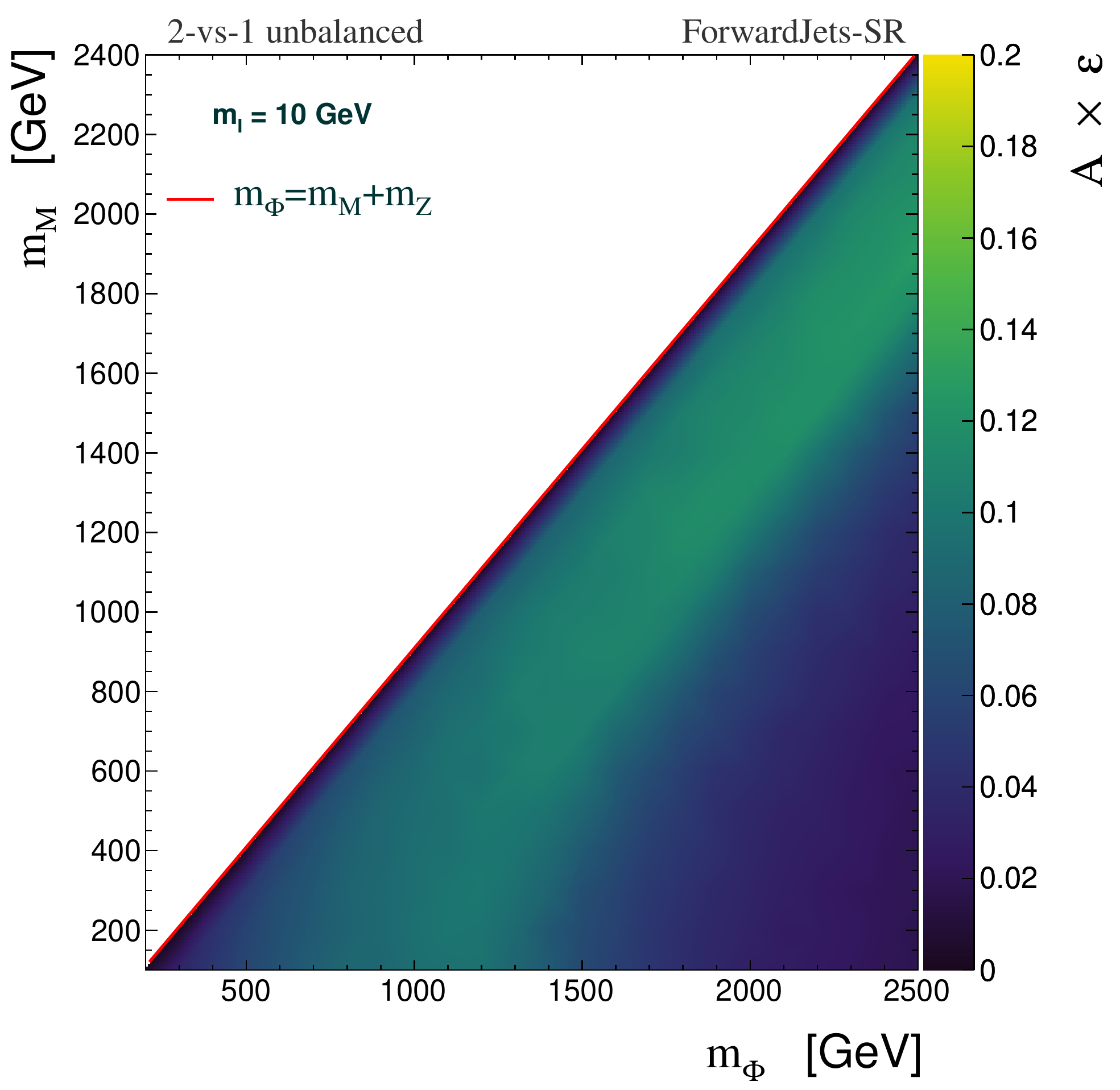}
    \hfill
    \includegraphics[width=0.49\textwidth]{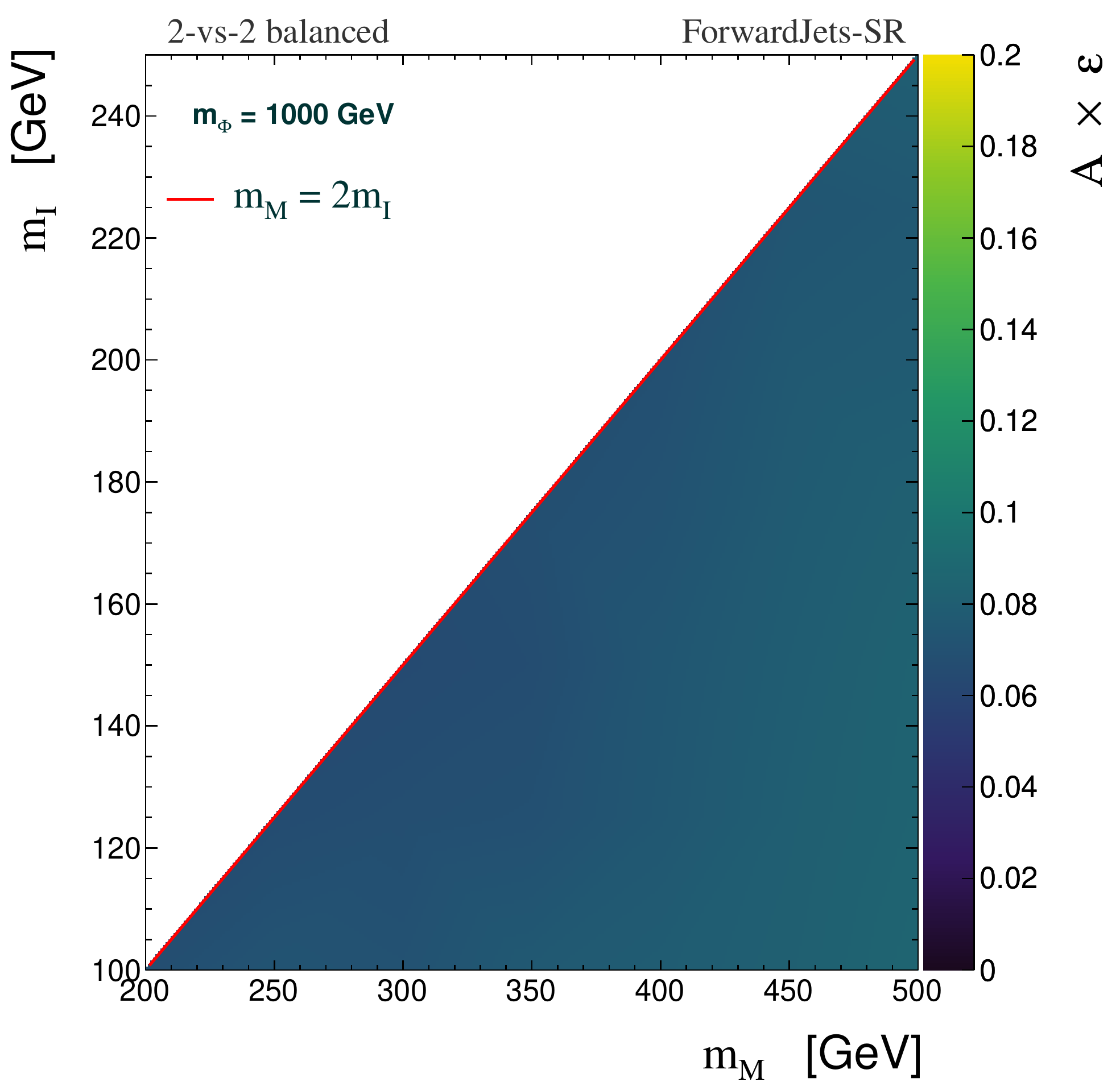}
    \caption{Analysis efficiency (including detector acceptance) for the selection corresponding to the ForwardJets-SR (for a sample produced with the charged lepton decays of the $Z$ boson). 
    The horizontal and vertical axes indicate the mass combination for which the efficiency is measured, while the color coding indicates the efficiency value obtained for this mass combination. The results are shown for the 1-vs-1 unbalanced (upper left panel), 2-vs-1 balanced (upper right panel), 2-vs-1 unbalanced (lower left panel), and 2-vs-2 balanced (lower right panel) topologies. The kinematic constraints for each topology are indicated by solid red lines, whereas the fixed mass value is specified in the legend.}
    \label{fig:Efficiency-ForwardJets-SR}
\end{figure}

In order to study the efficiency of the above selection for the four topologies described in \cref{sec:simp_model}, a scan on the mass parameters $(m_{\Phi},m_{M},m_{I})$ is carried out as specified in Tab.~\ref{tab:signal-scan}, and the overall efficiency is determined. For each topology one of the three mass parameters is kept fixed, while the other two masses are varied simultaneously in order to obtain a 2D map of the analysis efficiency per topology. This is performed individually for both signal regions defined above. The obtained values for the Standard-SR and the ForwardJets-SR are depicted in \cref{fig:Efficiency-Standard-SR} and \cref{fig:Efficiency-ForwardJets-SR} respectively.

First, one can notice that the Standard-SR has an overall larger efficiency for all signal topologies compared to the ForwardJets-SR, though the contribution of the latter is clearly non-negligible. The highest efficiency value for the Standard-SR occurs for the 2-vs-1 unbalanced topology and for a mass combination of about ($2100$ GeV, $1500$ GeV, $10$ GeV) reaching up to $0.168 \pm 0.002$\footnote{A value that is consistent within the statistical uncertainty is obtained for the 1-vs-1 unbalanced topology for ($2100$ GeV, $600$ GeV, $1500$ GeV).}, where the reported uncertainty is purely statistical. Similarly high values of the efficiency are also found for various points along the approximate line $m_{\Phi} - m_{M} = 600$ GeV for the 2-vs-1 unbalanced topology, and for points near $m_{\Phi} - m_{I} = 600$ GeV in the 1-vs-1 unbalanced topology. For the ForwardJets-SR, the highest efficiency is also reached for the 2-vs-1 unbalanced topology, but this time for a mass combination of ($2500$ GeV, $1900$ GeV, $10$ GeV), resulting in an efficiency of $0.128 \pm 0.002$. This demonstrates that the analysis is highly efficient for signal scenarios with a semi-boosted regime, which can be deduced from the event selection using relatively energetic leptons coming from the $Z$ boson. In the case of extremely boosted $Z$ bosons, the two lepton signatures overlap, impacting on their reconstruction and causing the isolation criteria to reject most of those events. An additional feature that is observed in the ForwardJets-SR for the two unbalanced topologies is that the efficiency gets larger for increasing mass $m_{\Phi}$ of the scalar resonance (for a fixed mass difference $m_\Phi-m_I$ (for 1-vs-1 unbalanced) or $m_\Phi-m_M$ (for 2-vs-1 unbalanced)). This is caused by the fact that the jets point more into the forward or backward direction as the mass of the heavy resonance is enhanced, which implies that they are more likely to pass the pseudorapidity difference cut included in the definition of the ForwardJets-SR. With regard to the variation in the signal kinematics for the various topologies, one can observe that the two unbalanced cases closely resemble each other, whereas the main qualitative difference occurs when comparing the balanced with the unbalanced cases. For the balanced topologies, the variation in the selection efficiency is less pronounced as one moves across the scanned phase space, given that here $m_{\Phi}$ has been fixed. The most visible variation in the 2-vs-1 balanced topology occurs (besides kinematic edges) where the difference $m_{M}-m_{I}$ is increased, which results in a slightly more boosted $Z$ boson and a more sizable imbalance between the $Z$ boson and the \pTmissVec. In the 2-vs-2 balanced topology, the variations in $m_{M}-m_{I}$ can barely modify the above-mentioned imbalance, since an invisible decay of the mediator occurs in both legs of the diagram.

%%%%%%%%%%%%%%%%%%%%%%%%%%%%%%%%%%%%%%%%%%%%%%%%%%%%%%%%%%%%%%%%%%%%%%%%%%%%%%

\subsection{Background expectation}
\label{sec:background-expectation}

The multiple background processes contributing to the selection described in \cref{sec:event-selection} are estimated using simulation, which is detailed in \cref{sec:mc_simulation}. The main SM processes that pass the selection are $Z+\text{jets}$, $t\bar{t}$, single-top ($tW$ channel), and di-boson production. The $Z+\text{jets}$ process enters in the Standard-SR selection if a relatively boosted $Z$ boson is produced in association with a $b$-quark, whereas if a similarly energetic $Z$ boson is produced through di-boson production it becomes more likely that it will contribute to the ForwardJets-SR. The $t\bar{t}$ production with di-leptonic decay also poses a substantial background component, given the presence of $b$-quarks in the final state and the fact that \pTmiss is generated by neutrinos. The production of a single top quark, specifically in the $tW$ channel with di-leptonic decay, exhibits a very similar signature compared to $t\bar{t}$, only differentiated at Born level by having one less $b$-quark. In both cases, if the $b$-quark(s) are not tagged as $b$-jet(s) during the selection, the event could end up in the ForwardJets-SR, with possible contributions of other jets arising from QCD initial or final state radiation. $WW$, $WZ$, and $ZZ$ production with additional jets can also enter both signal regions. For $WZ$, if the lepton from the $W$ boson is not reconstructed, this background resembles the signal signature; this also happens if one of the $Z$ bosons in the $ZZ$ process decays into a pair of neutrinos while the other decays into a lepton pair. In the case of the $WW$ process, the signature is similar to that of the $t\bar{t}$ and $tW$ processes, where additional ($b$-)jets can arise from QCD radiation.

A summary of the expected event yields for the various background components assuming an integrated  luminosity of 137 fb\textsuperscript{-1} \footnote{This is approximately the amount of luminosity collected by the CMS experiment during the Run 2 data taking period (see~\ccite{Run2:Luminosity}), and it will be used as reference for most of the results presented in this work.} is reported in Tab.~\ref{tab:background-yields}.
\begin{table}[h]
  \begin{center}
    \renewcommand{\arraystretch}{1.}
    \begin{tabular}{|l||c|c|}
      \hline
      Process & Standard-SR & ForwardJets-SR \\ 
      \hline\hline
      $Z+\text{jets}$ & $350147 \pm 18557$ & $618659 \pm 24667$ \\
      $t\bar{t}$ & $72839 \pm 421$\phantom{0} & $7672 \pm 137$ \\
      Single-top ($tW$) & $6380 \pm 59$\phantom{0} & $826 \pm 21$ \\
      $WW$ & \phantom{0}$77 \pm 13$ & $224 \pm 22$ \\
      $WZ$ & $143 \pm 6$\phantom{0} & $749 \pm 14$ \\
      $ZZ$ & $149 \pm 3$\phantom{0} & $462 \pm 5$\phantom{0} \\
      \hline
    \end{tabular}
    \caption{ Yield estimates in the two defined signal regions for the different background processes contributing after the selection specified in \cref{sec:event-selection} for a reference integrated luminosity of 137 fb\textsuperscript{-1}. The reported uncertainty represents solely the statistical component. }
    \label{tab:background-yields}
  \end{center}
\end{table}

As can be noticed, the remaining background after the selection in both signal regions is relatively high, and this can be understood if one takes into account that no requirement has been imposed on \pTmiss. This has been done on purpose to keep the selection as inclusive as possible with respect to highly diversified kinematics resulting from the multiple signal scenarios explored; the strategy to follow in order to boost the sensitivity for all signal topologies will be outlined in the next section. \cref{tab:background-yields} elucidates the fact that $Z+\text{jets}$ amply dominates the background composition in both signal regions, though this particular background will be relatively easy to discriminate using \pTmiss, as it will be illustrated in \cref{sec:limit-extraction-method}. Another clear observation extracted from the above table is that $t\bar{t}$ and $tW$ backgrounds are more prominent in the Standard-SR, something that would be expected given the $b$-jet requirement in this region and the $b$-jet veto in the ForwardJets-SR. The consequences of that will be more clear when evaluating the expected sensitivity of the analysis per signal region, as despite the analysis having a larger efficiency in the Standard-SR (see \cref{fig:Efficiency-Standard-SR} and \cref{fig:Efficiency-ForwardJets-SR}), the smaller contribution of processes with a substantial tail in the \pTmiss distribution ($t\bar{t}$ and $tW$) provides the ForwardJets-SR with a solid discriminating power.

%%%%%%%%%%%%%%%%%%%%%%%%%%%%%%%%%%%%%%%%%%%%%%%%%%%%%%%%%%%%%%%%%%%%%%%%%%%%%%

\subsection{Limit extraction method}
\label{sec:limit-extraction-method}

As mentioned in the previous section, an optimisation of the analysis for a specific signal topology will lead to a drastic reduction in sensitivity for other kinematically distinct topologies. Consequently, the strategy devised for the present analysis was to keep the selection as inclusive as possible and to try to implement a robust statistical analysis in one of the most powerful observables to achieve a strong level of discrimination in the majority of signal regimes. The variable designated to accomplish that task was the \pTmiss, given the huge discrimination against the dominant $Z+\text{jets}$ background (which does not have a genuine \pTmiss\ contribution) and the moderate separation against all the other backgrounds mentioned in \cref{sec:background-expectation}. The statistical inference is then performed using a shape scanning over the binned \pTmiss distribution. The selected binning, which has been chosen according to both statistical and sensitivity considerations, is shown in \cref{tab:met-binning} for both signal regions.

\begin{table}[h]
  \begin{center}
    \renewcommand{\arraystretch}{1.}
    \begin{tabular}{|l||c|c|}
      \hline
      Region & Number of \pTmiss bins & Bin edges in \pTmiss [GeV] \\ 
      \hline\hline
      Standard-SR & 10 & \multicolumn{1}{l|}{$[0,20,40,70,100,135,190,280,400,550,1000]$} \\
      ForwardJets-SR & 8 & \multicolumn{1}{l|}{$[0,20,40,70,100,150,240,360,1000]$} \\
      \hline
    \end{tabular}
    \caption{ Binning in \pTmiss chosen for each of the two signal regions. All events with $\pTmiss > 1000\text{ GeV}$ are added to the last bin of the respective region.}
    \label{tab:met-binning}
  \end{center}
\end{table}

\begin{figure}
    \centering
    \includegraphics[width=0.49\textwidth]{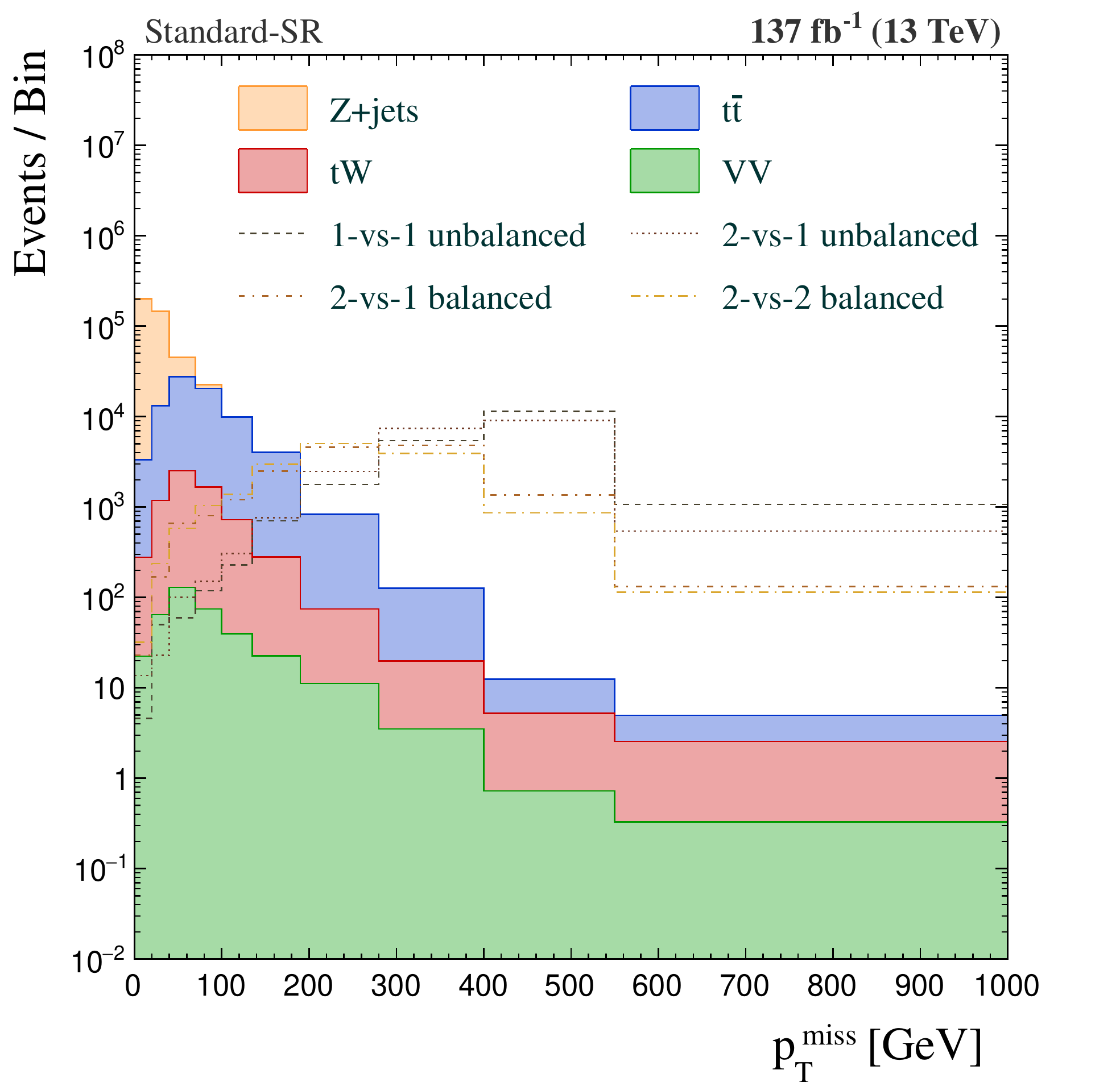}
    \hfill
    \includegraphics[width=0.49\textwidth]{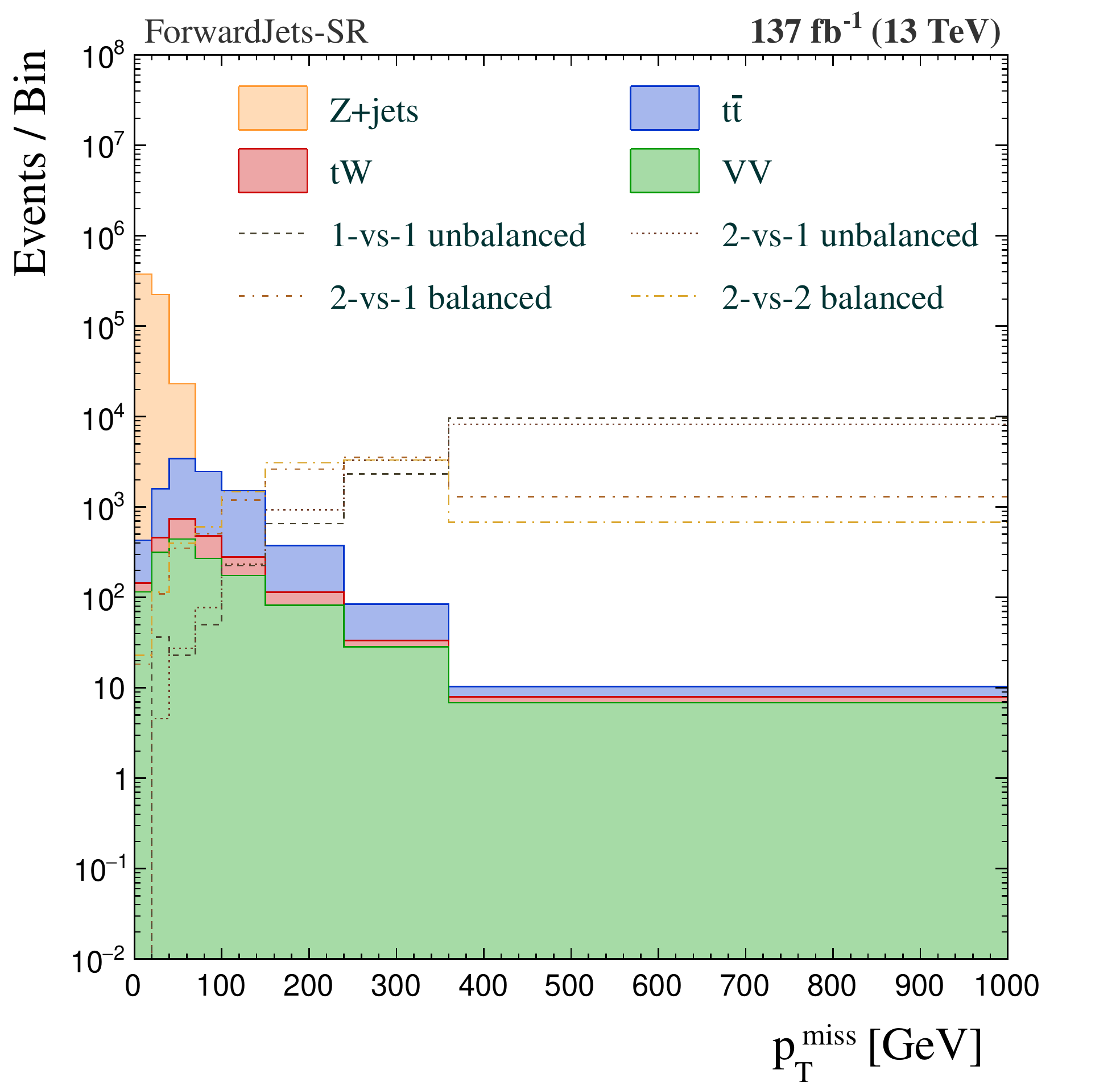}
    \caption{
    Binned \pTmiss distributions for both signal regions, Standard-SR (left panel) and ForwardJets-SR (right panel), constructed to perform the statistical analysis corresponding to the limit extraction procedure. All processes, including the benchmarks chosen for the signal topologies, have been normalized to an integrated luminosity of 137 fb\textsuperscript{-1}, where for all the signal benchmark scenarios shown (dashed histograms), a common cross section of 1 pb has been used for their normalization. For all signal topologies, the mass combination used was ($1000$ GeV, $300$ GeV, $100$ GeV), except for the 1-vs-1 unbalanced topology, where $m_{M}$ is irrelevant (see above),
    and the 2-vs-1 unbalanced topology, where $m_{I}=10$ GeV was used; it should be noted that these modifications do not significantly alter the shape of the \pTmiss distribution. The various background processes have been stacked up to illustrate the overall background contribution. The various di-boson processes ($WW$, $WZ$, and $ZZ$) have been grouped under the label $VV$.}
    \label{fig:Met-distributions}
\end{figure}

The corresponding \pTmiss distributions for all background processes and for a few benchmark signal points are presented in \cref{fig:Met-distributions}. It can be noticed there that the $Z+\text{jets}$ process has a negligible contribution in bins with high signal expectation, therefore weakly impacting the full significance of the analysis. The other processes, however, do present relatively large \pTmiss due to the genuine presence of invisible particles, and are the contributions that determine in the end the overall analysis sensitivity. In terms of signal shapes, one can see a clear difference between the balanced and the unbalanced cases, but among signal topologies of the same kind (e.g.\ 1-vs-1 unbalanced compared to 2-vs-1 unbalanced), the distinction is much less pronounced, in agreement with what was perceived from the signal efficiency maps obtained in \cref{sec:efficiency-maps}.

The statistical model is built using the template shapes displayed in Fig.~\ref{fig:Met-distributions} for each given process. On top of the statistical uncertainty resulting from the limited number of MC events in the samples per bin, various systematic uncertainties affecting the normalization have been incorporated to make the results more realistic. An uncertainty has been assigned to the integrated luminosity value used, corresponding to 2.5\%, where we refer to the measured uncertainty value by CMS in 2018~\cite{CMS:2019jhq}. An uncertainty on the normalization of the $t\bar{t}$ process, and consequently affecting only this template, with a value of 5\% is included, assuming the latest precision achieved by the theoretical calculations of the cross section~\cite{Czakon:2013goa}. Similarly, a value of also 5\%~\cite{Kidonakis:2013zqa} is considered for the single-top production through the $tW$ channel. For the case of the $Z+\text{jets}$ process, and given the more specific configuration used to generate events, we have used the uncertainty value obtained after running the FEWZ 3.1 program presented in~\ccite{Li:2012wna}, which corresponds to a total uncertainty of 2.5\% on the cross section. For $VV$, and taking into account the combination of multiple individual processes into this template, a conservative 15\% uncertainty is assumed, which is close to the uncertainty value delivered by most NLO MC generators like MC@NLO and POWHEG for these processes.

The limits are computed using the \textsc{RooStats} package~\cite{RooStats:Twiki} by means of the \textsc{HistFactory} interface~\cite{Cranmer:1456844} that is implemented in ROOT~\cite{Antcheva:2009zz}. The calculation is made using the asymptotic approximation in \textsc{RooStats}, which is an implementation of the results obtained in~\ccite{Cowan:2010js}, and also considering the CL\textsubscript{s} method described in~\ccite{Read:451614}. Two categories corresponding to the two signal regions are constructed. Thus, limits for each individual category as well as for the combination of both categories are provided. A confidence level (CL) of 95\% has been used to compute all the limits that will be presented in this work.

%%%%%%%%%%%%%%%%%%%%%%%%%%%%%%%%%%%%%%%%%%%%%%%%%%%%%%%%%%%%%%%%%%%%%%%%%%%%%%

%%%%%%%%%%%%%%%%%%%%%%%%%%%%%%%%%%%%%%%%%%%%%%%%%%%%%%%%%%%%%%%%%%%%%%%%%%%%%%
%%%%%%%%%%%%%%%%%%%%%%%%%%%%%%%%%%%%%%%%%%%%%%%%%%%%%%%%%%%%%%%%%%%%%%%%%%%%%%

\section{Expected sensitivity}
\label{sec:results}

%%%%%%%%%%%%%%%%%%%%%%%%%%%%%%%%%%%%%%%%%%%%%%%%%%%%%%%%%%%%%%%%%%%%%%%%%%%%%%

 The expected sensitivity of this analysis is evaluated by setting limits in the absence of a signal on the product of the production cross section times the full branching fraction for a given topology, i.e.\ $\sigma(b\bar{b}\Phi)\times \mathcal{B}(\Phi \rightarrow Z + \pTmiss) \times \mathcal{B}(Z \rightarrow l\bar{l})$. Here, the term $\mathcal{B}(\Phi \rightarrow Z + \pTmiss)$ depends on the specific decay case covered by the signal topology; as an example, for the 2-vs-1 unbalanced topology (Fig.~\ref{fig:monoZ_topologies_2vs1_unbalanced}) that terms read $\mathcal{B}(\Phi \rightarrow Z + \pTmiss) \equiv \mathcal{B}(\Phi \rightarrow Z M) \times \mathcal{B}(M \rightarrow II)$.

Two sets of results are presented in terms of limits on the cross section. The first group corresponds to 1D projections, where two of the mass parameters are fixed and the remaining mass is varied. In those results the sensitivity of the two defined signal regions can easily be compared with each other and with the combined sensitivity. Furthermore, we will indicate the bands corresponding to one and two standard deviations from the reported central value. The second set of results are 2D limit scans corresponding to the ones presented in \cref{fig:Efficiency-Standard-SR,fig:Efficiency-ForwardJets-SR} for the efficiency case. In those results we display the central limit value obtained from the combination of the two signal regions.

%%%%%%%%%%%%%%%%%%%%%%%%%%%%%%%%%%%%%%%%%%%%%%%%%%%%%%%%%%%%%%%%%%%%%%%%%%%%%%
\subsection{Limit Scan in 1D}
\label{sec:result-limit-1D}

\begin{figure}
    \centering
    \includegraphics[width=0.49\textwidth]{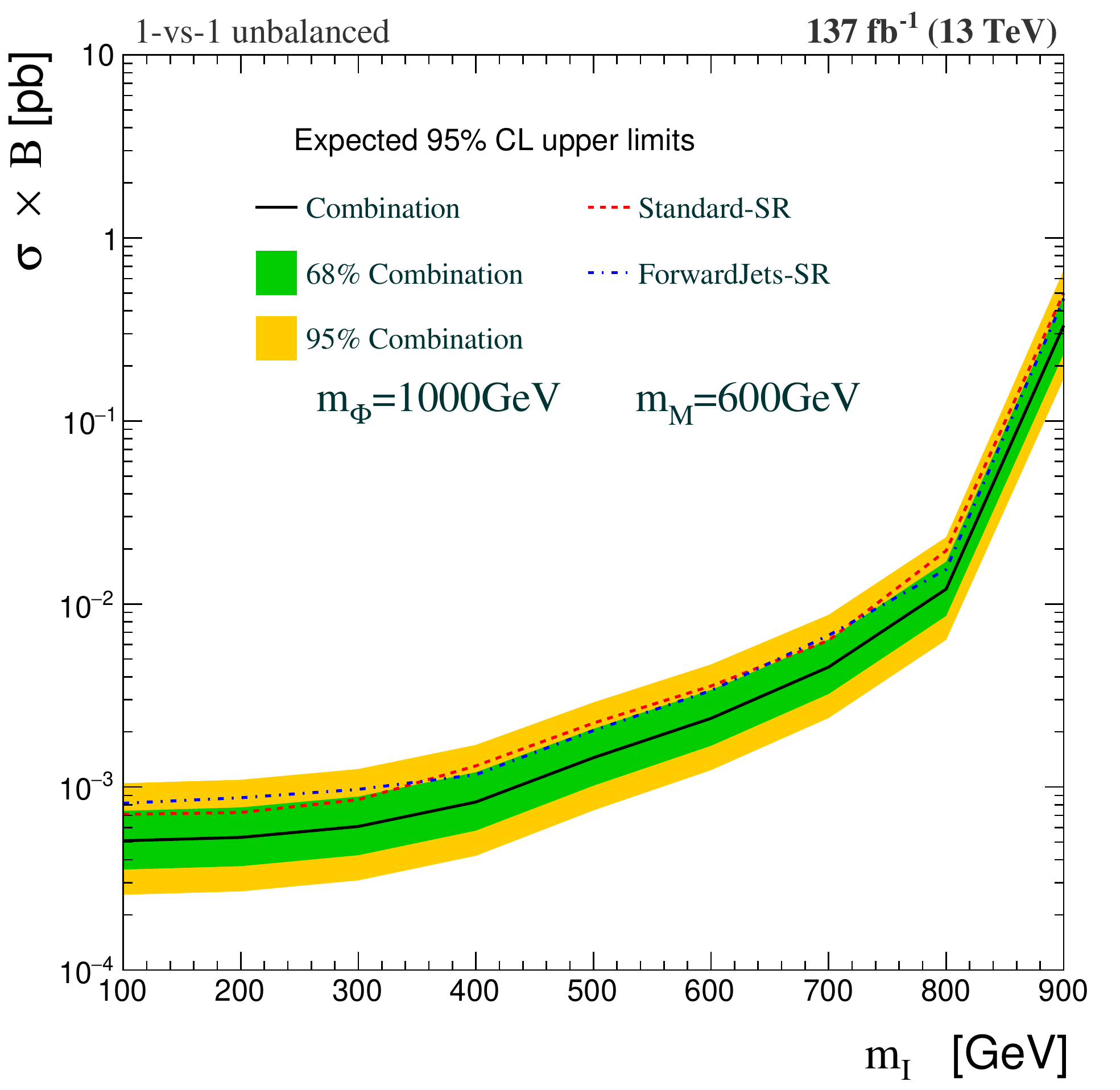}
    \includegraphics[width=0.49\textwidth]{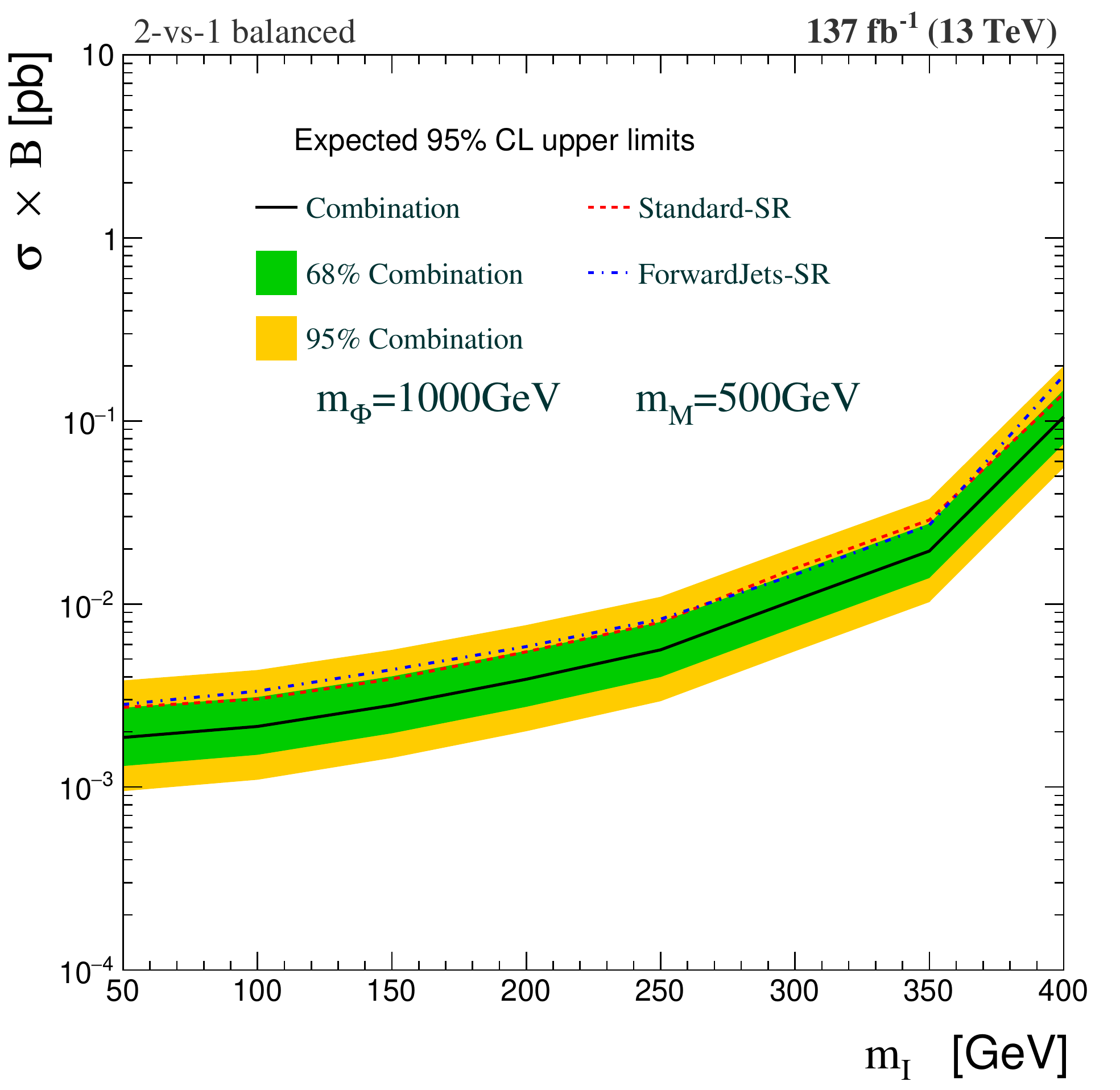}
    \includegraphics[width=0.49\textwidth]{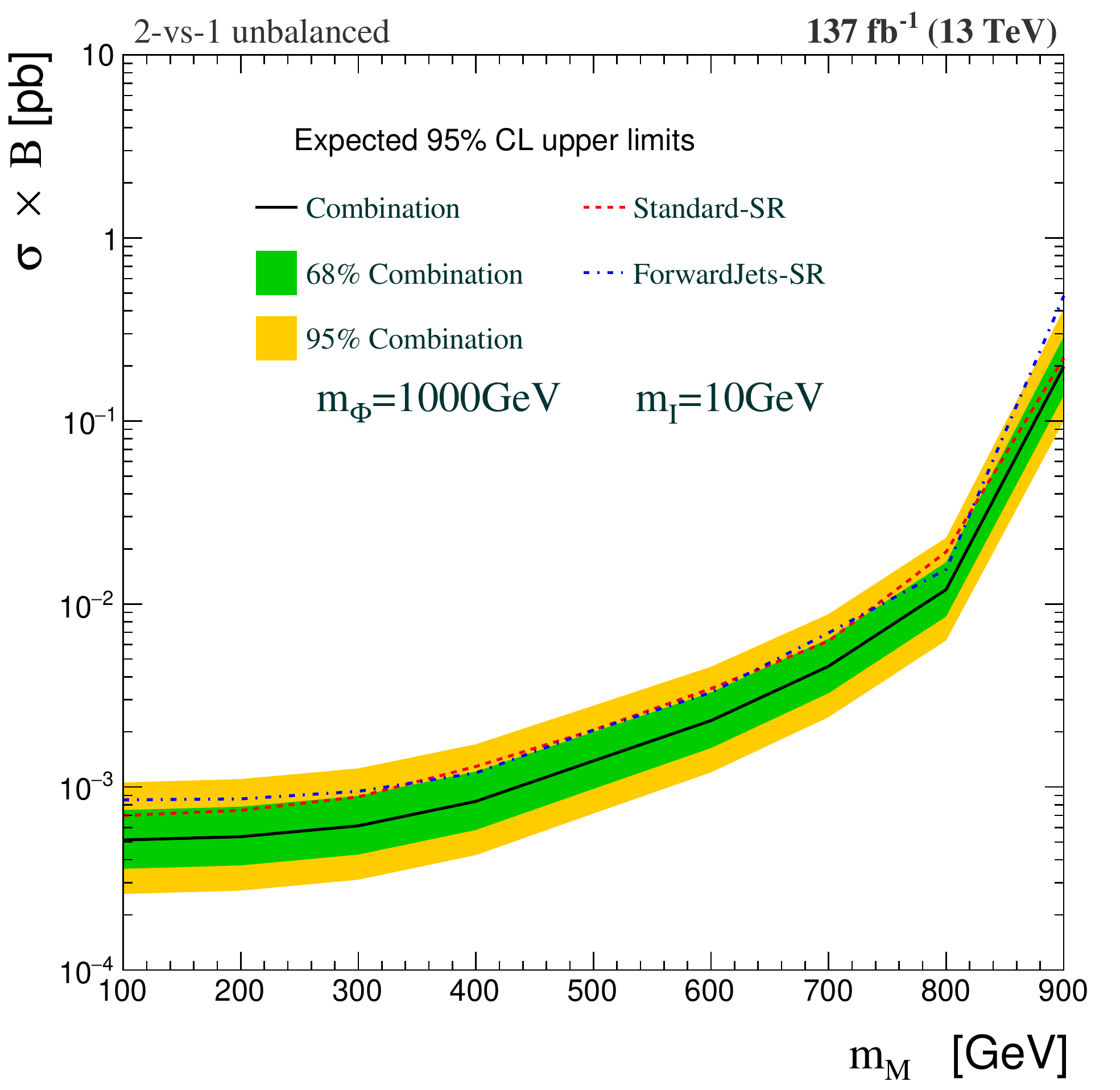}
    \includegraphics[width=0.49\textwidth]{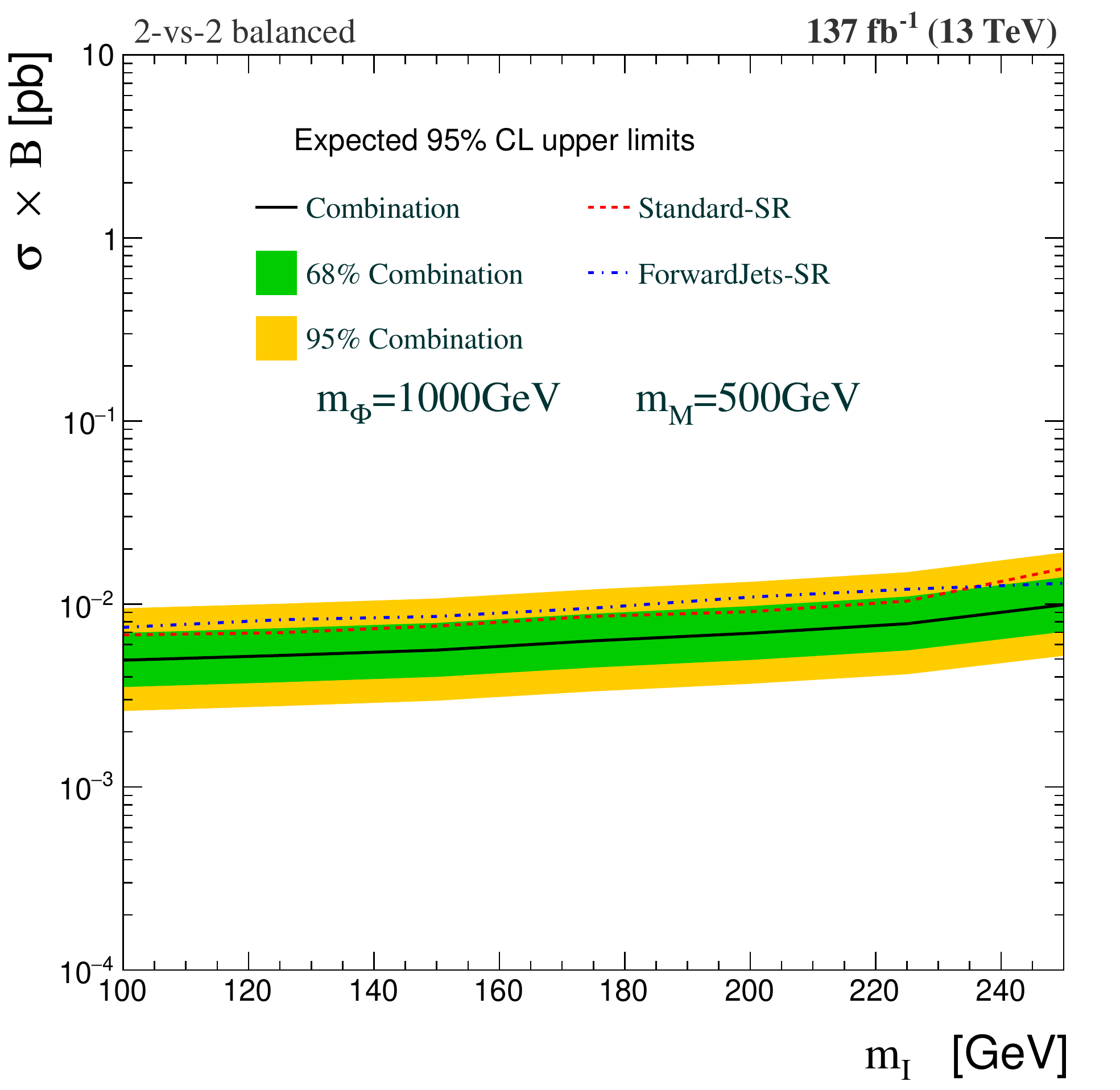}
    \caption{Expected upper limits on the cross section times branching fraction (as defined in the text) for a 1D variation of one of the mass parameters in each topology. The horizontal axis indicates the varied mass parameter, while the vertical axis represents the upper limit obtained at 95\% CL for various scenarios: the central limit obtained using only the Standard-SR (red dotted line), the central limit obtained using only the ForwardJets-SR (blue dotted line), and the central limit obtained combining the two signal regions (black solid line). The green and yellow uncertainty bands correspond to the 68\% and 95\% interval coverage for the combined limit, respectively. The results are shown for the 1-vs-1 unbalanced (upper left panel), 2-vs-1 balanced (upper right panel), 2-vs-1 unbalanced (lower left panel), and 2-vs-2 balanced (lower right panel) topologies. The choice for the two mass parameters that have been kept fixed is indicated in the legend for each signal topology.}
    \label{fig:Limits_1D_v1}
\end{figure}

\begin{figure}
    \centering
    \includegraphics[width=0.49\textwidth]{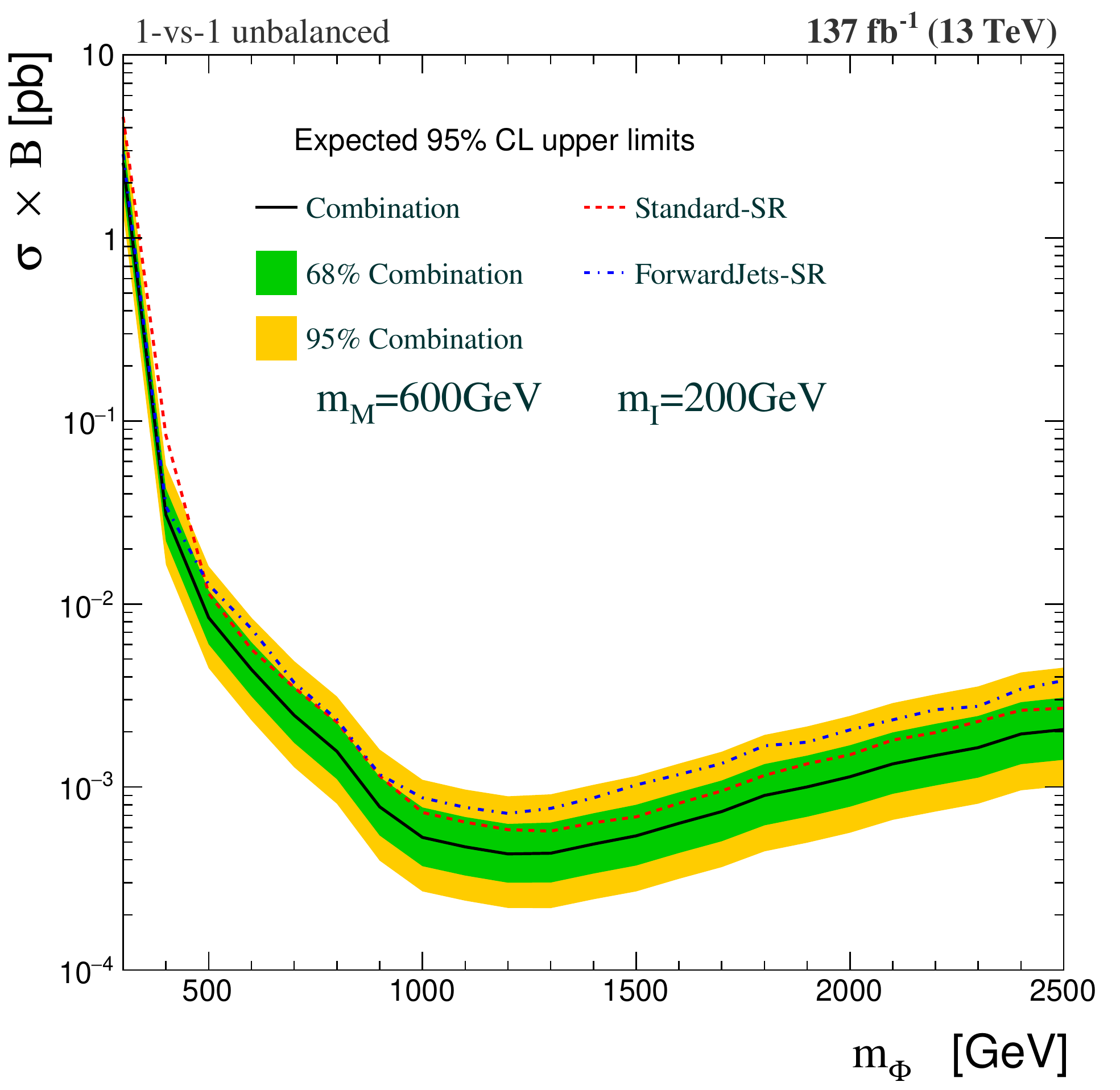}
    \includegraphics[width=0.49\textwidth]{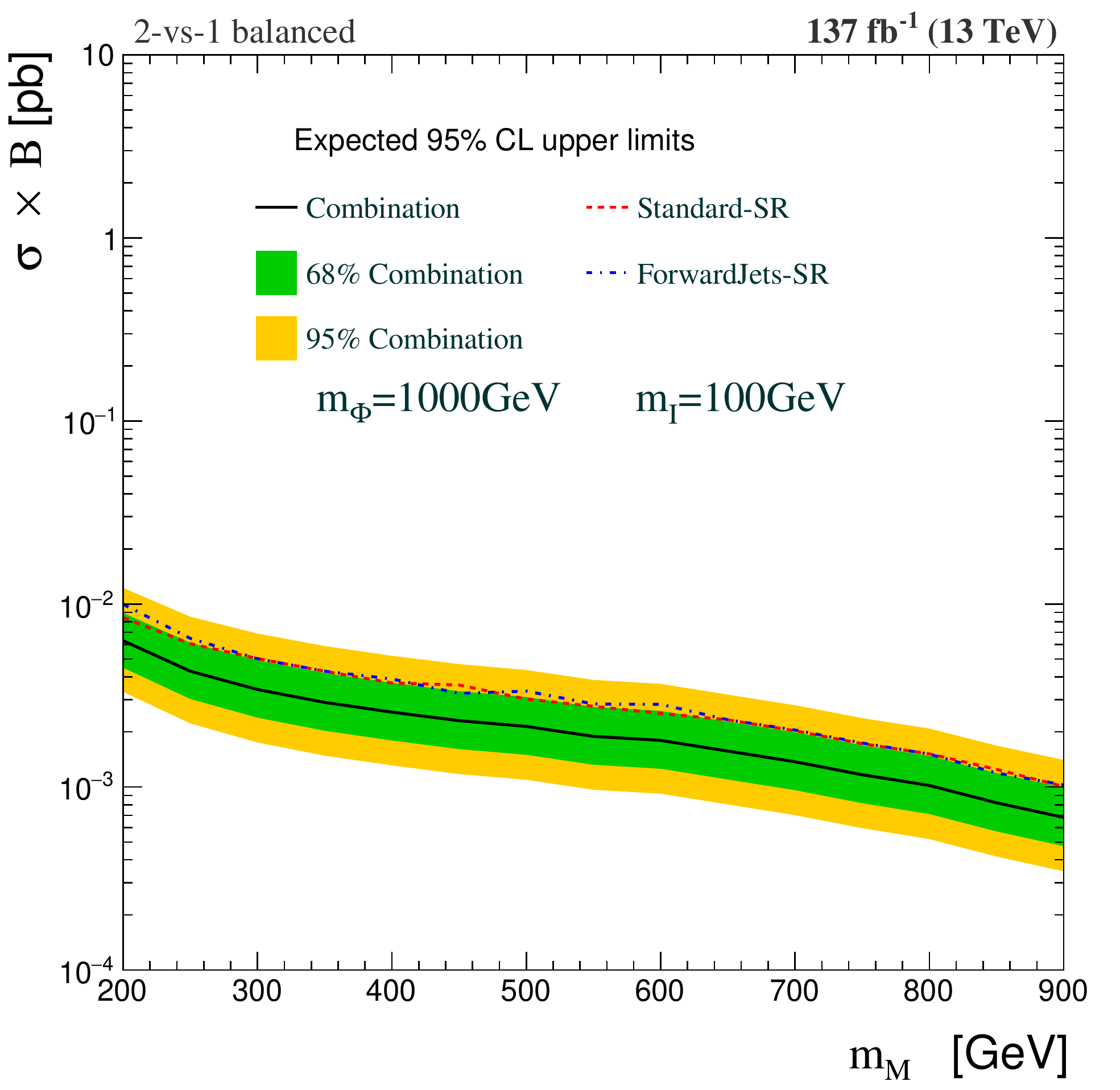}
    \includegraphics[width=0.49\textwidth]{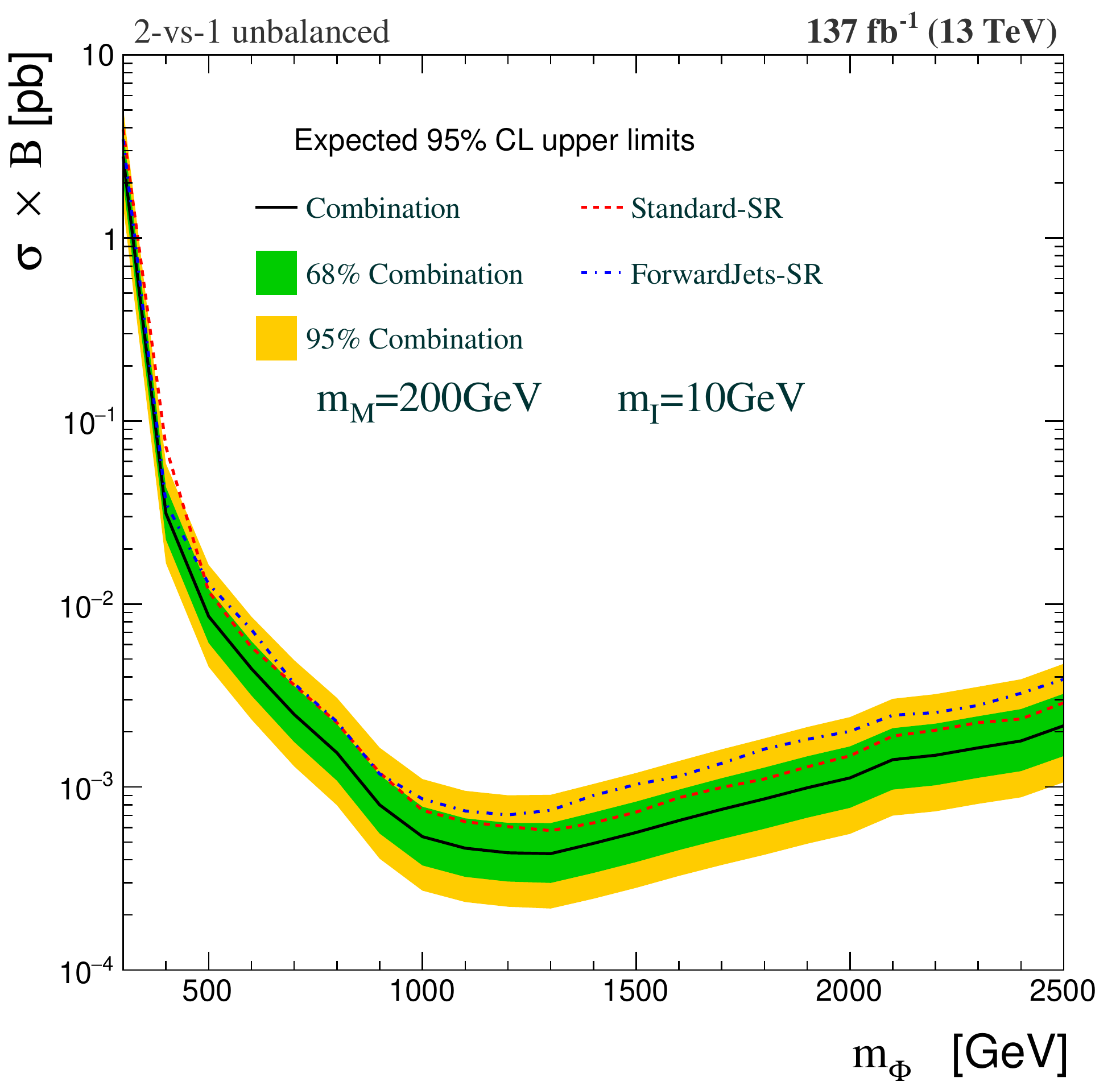}
    \includegraphics[width=0.49\textwidth]{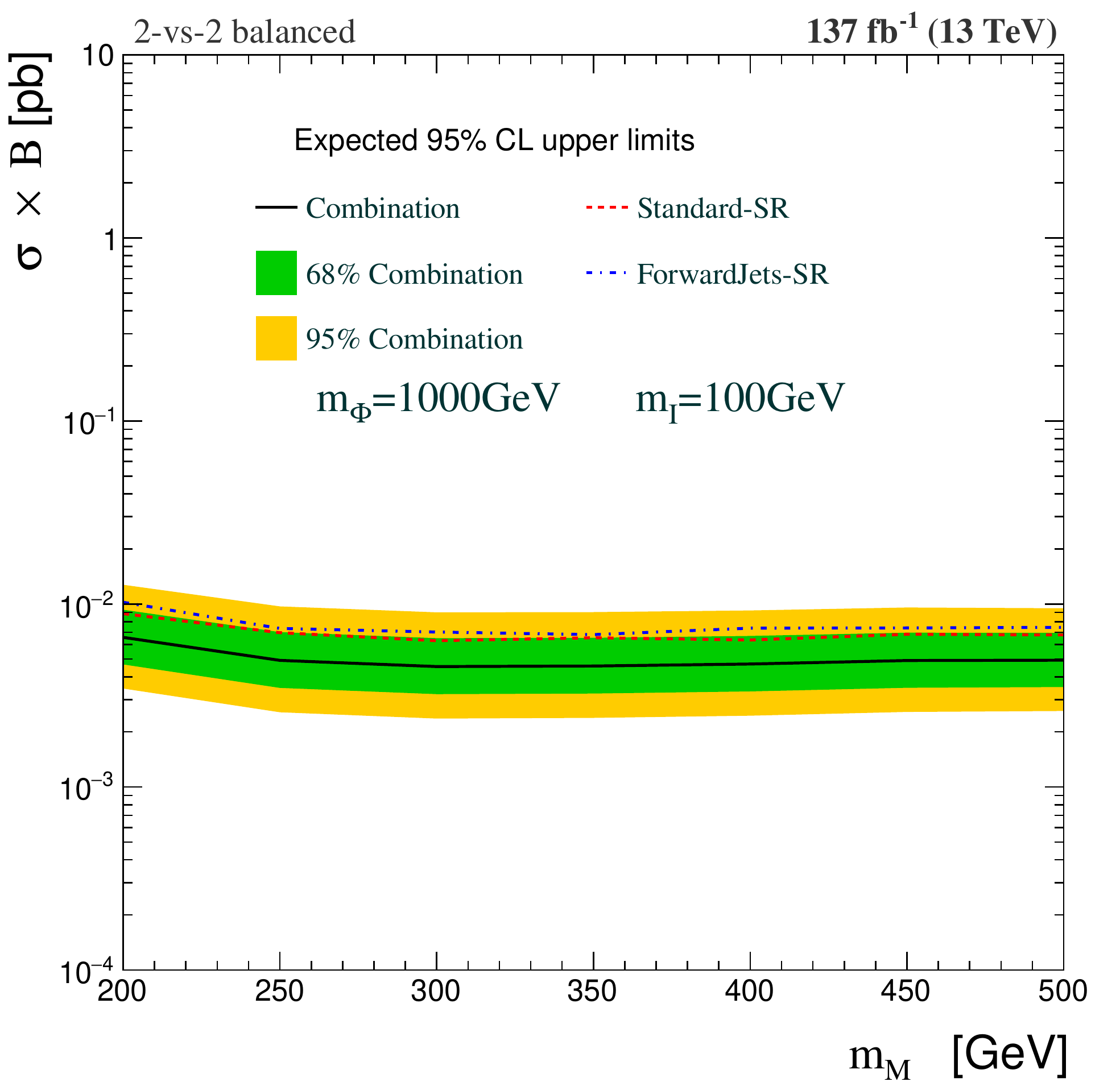}
    \caption{As in \cref{fig:Limits_1D_v1}, but for different choices of the varied and fixed mass parameters.}
    \label{fig:Limits_1D_v2}
\end{figure}

The 1D limit projections are shown in \cref{fig:Limits_1D_v1,fig:Limits_1D_v2}. The first observation to be made is with respect to the individual sensitivity reached by the two signal regions. The individual limits are close to each other for all different variations presented in the figures, which points to the fact that the two regions are similarly important with respect to the overall sensitivity of the analysis. This is explained by the limited efficiency of identifying one $b$-tagged jet for the signal scenarios that are investigated here, caused by both the forward kinematics of the jets and the performance of the $b$-tagging algorithm. On the other hand, the low rate of forward jets in the background processes dominating for this di-lepton selection makes the ForwardJets-SR highly competitive despite having a slightly lower overall signal efficiency, as this signal region benefits from a drastic reduction of the overwhelming high-\pTmiss top background. This becomes particularly evident in the $m_{I}$ and $m_{M}$ scans for the unbalanced topologies in \cref{fig:Limits_1D_v1}, where one can see that the ForwardJets-SR provides almost identical sensitivity to the Standard-SR in the entire mass range.

Another interesting piece of information that can be extracted from these results, and that was also asserted in the analysis of the efficiencies in Sec.~\ref{sec:efficiency-maps}, is that there is no intrinsic difference in the signal kinematics for the two unbalanced topologies; in good approximation these two topologies can be treated such that they yield identical results, and this is verified in both projections shown in \cref{fig:Limits_1D_v1,fig:Limits_1D_v2}. This means that any experimental analysis optimized for one of those two topologies is automatically optimal for the other one as well. In the case of the analysis presented in this work, it can be corroborated from the limits that our analysis is highly sensitive to scenarios with semi-boosted $Z$ bosons, occurring for a mass difference $m_{\Phi}-m_I$ (for 1-vs-1 unbalanced) or $m_{\Phi}-m_M$ (for 2-vs-1 unbalanced) of around 1 TeV. The situation for the balanced topologies is slightly more involved. Even though the results are very close to each other if $m_{M}$ is sufficiently small (see Fig.~\ref{fig:Limits_1D_v2} right panel), with increasing mediator mass the limits depart from each other substantially, and the limits for the 2-vs-1 balanced topology are found to be up to a factor of two stronger compared to the 2-vs-2 balanced topology for the same mass parameters. This effect can be explained from the discussion of the efficiency results in Sec.~\ref{sec:efficiency-maps}. It arises from the fact that the $Z$ boson becomes more boosted for the 2-vs-1 balanced case if the difference $m_{M}-m_{I}$ increases.

%%%%%%%%%%%%%%%%%%%%%%%%%%%%%%%%%%%%%%%%%%%%%%%%%%%%%%%%%%%%%%%%%%%%%%%%%%%%%%
\subsection{Limit Scan in 2D}
\label{sec:result-limit-2D}

\begin{figure}
    \centering
    \includegraphics[width=0.49\textwidth]{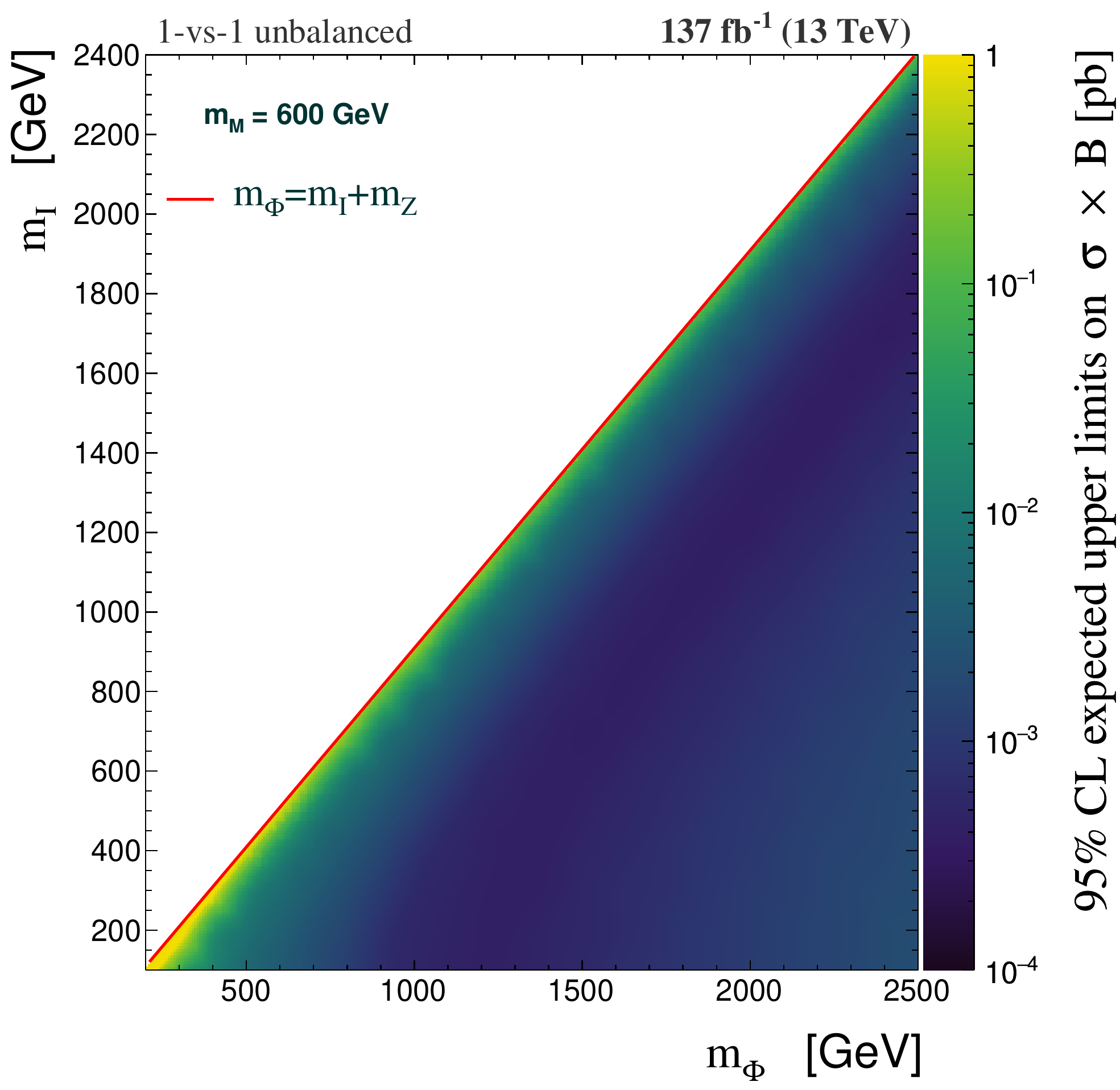}
    \includegraphics[width=0.49\textwidth]{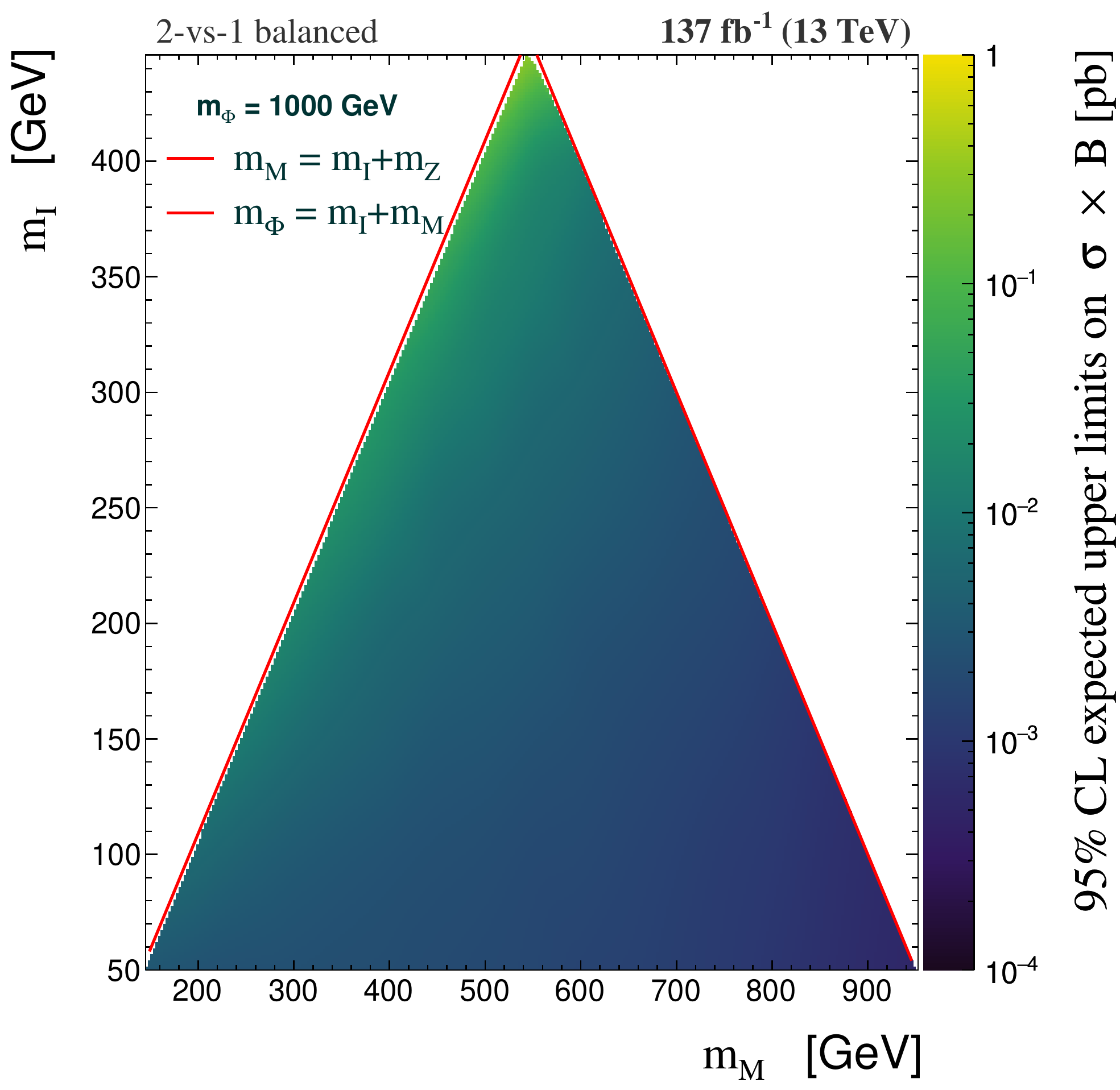}
    \includegraphics[width=0.49\textwidth]{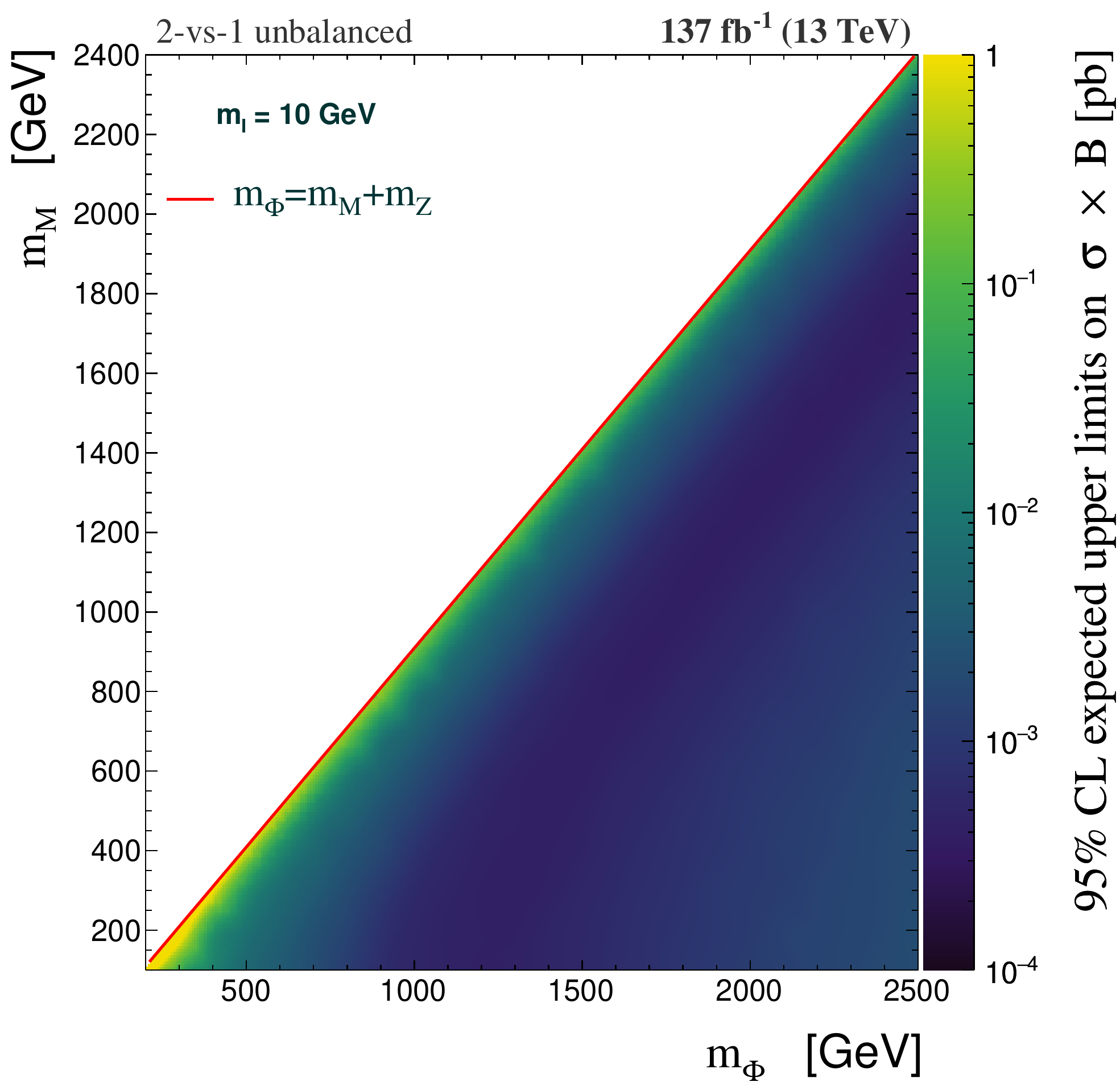}
    \includegraphics[width=0.49\textwidth]{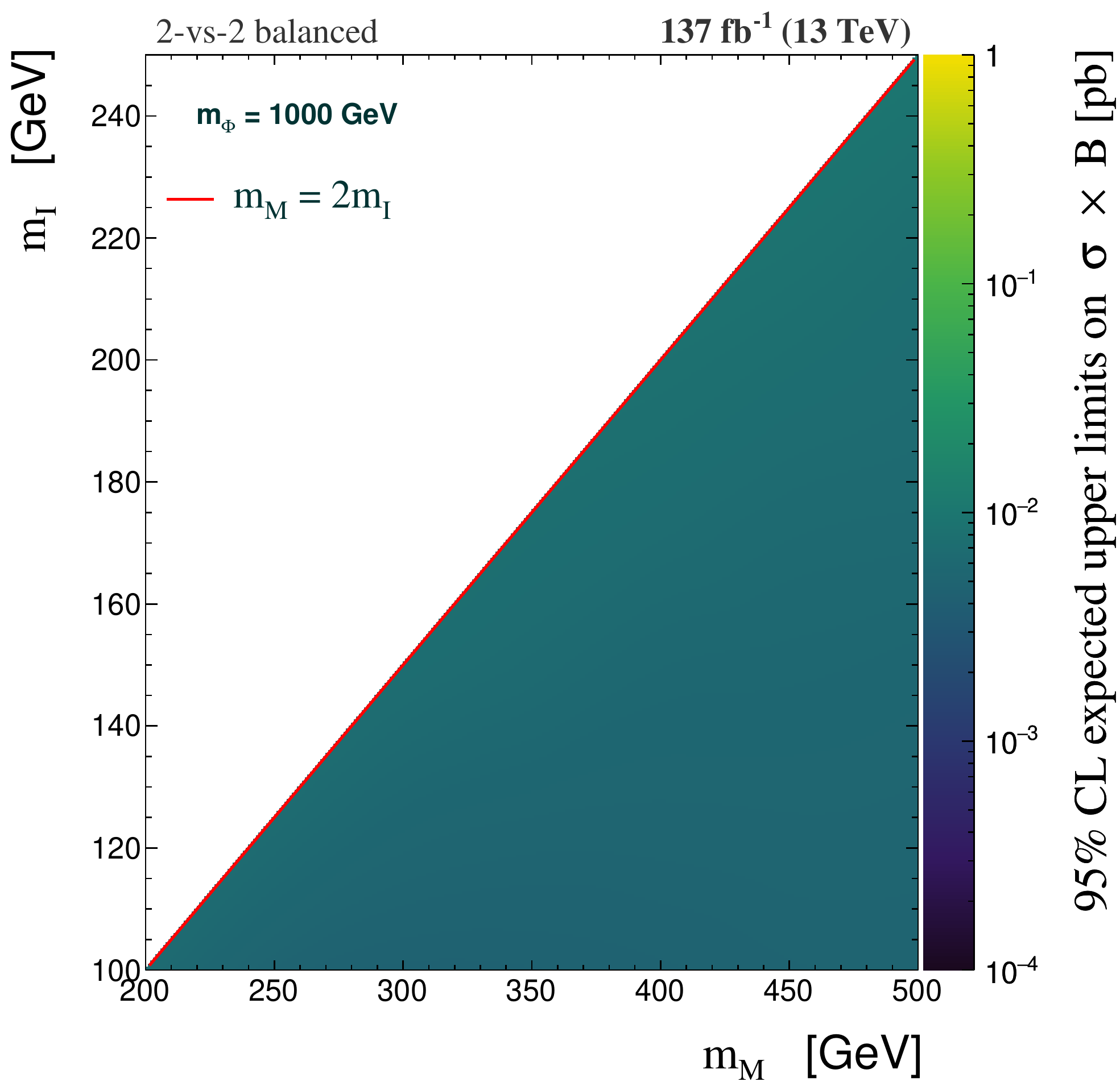}
    \caption{Expected upper limits on the cross section times branching fraction (as defined in the text) for a 2D variation of the mass parameters in each topology. The horizontal and vertical axes indicate the mass combination for which the limit is calculated, while the color coding represents the central value obtained at 95\% CL combining the two signal regions for that specific 2D point. The results are shown for the 1-vs-1 unbalanced (upper left panel), 2-vs-1 balanced (upper right panel), 2-vs-1 unbalanced (lower left panel), and 2-vs-2 balanced (lower right panel) topologies. The kinematic constraints for each topology are indicated by solid red lines, whereas the fixed mass value is specified in the legend.}
    \label{fig:Limits_2D}
\end{figure}

Analogously as it was done for the signal efficiency in \cref{sec:efficiency-maps}, a scan in the 2D plane formed by the two mass parameters that are varied is performed to get the cross section limit map for each signal topology. The results are shown in \cref{fig:Limits_2D}, where the limit obtained from the combination of the two signal regions is depicted. 

At first glance, a comparison with the patterns observed in the efficiency maps reveals a quite close similarity between the upper limits and the signal efficiency. This effect is of course expected given that the signal normalization is the main factor contributing to the constraint imposed on the cross section. Nevertheless, a closer look at \cref{fig:Limits_2D}, in particular at the limit maps corresponding to the unbalanced topologies (left panel), evidences a shift of the region with the most stringent limits with respect to the region with the largest efficiency towards higher mass differences $m_{\Phi}-m_I$ (for 1-vs-1 unbalanced) or $m_{\Phi}-m_M$ (for 2-vs-1 unbalanced). The size of this shift for the unbalanced cases ranges from 100 GeV to 400 GeV depending on the heavy scalar mass, and is a consequence of exploiting the shape of the \pTmiss\ distribution as a further constraint on the models. For larger mass differences (above $\sim 600\text{ GeV}$) the efficiency tends to drop due to the collimation of the leptons from the $Z$ boson decay, but at the same time the \pTmiss distribution starts to peak at higher values, which compensates the loss in signal acceptance by an increased significance in the last bins of the \pTmiss distribution (see \cref{fig:Met-distributions}). This effect is less obvious to discern for the two balanced topologies where the additional boosting created by more unbalanced configurations is either almost non-existent (2-vs-2 balanced) or localized in the corners of the phase space (2-vs-1 balanced). The strongest upper limits obtained for the 1-vs-1 unbalanced and the 2-vs-1 unbalanced topologies are around $3.5 \times 10^{-4}~\text{pb}$, whereas they are close to $5.5 \times 10^{-4}~\text{pb}$ and $4.5 \times 10^{-3}~\text{pb}$ for the 2-vs-1 balanced and 2-vs-2 balanced topologies, respectively. In terms of similarities among the signal topologies, \cref{fig:Limits_2D} confirms once again that the unbalanced topologies lead to an equivalent kinematic configuration, while the two balanced scenarios lead to a slight difference in kinematics if they are compared to each other for the same mass range. Our results presented here and in \cref{sec:result-limit-1D} show that a clear distinction can be made between the unbalanced and the balanced scenarios, which motivate specific experimental analyses optimized for each case individually.

%%%%%%%%%%%%%%%%%%%%%%%%%%%%%%%%%%%%%%%%%%%%%%%%%%%%%%%%%%%%%%%%%%%%%%%%%%%%%%

\subsection{2HDMa interpretation}

After deriving model-independent limits, we here demonstrate how these limits can be employed to constrain a concrete BSM model. As an exemplary model,\footnote{See \ccite{Bahl:2021xmy} for more examples of models with similar experimental signatures.} we study the Two-Higgs-Doublet model (2HDM) with an additional pseudoscalar, which serves as a dark matter portal. This model is usually referred to as 2HDMa. 

In addition to the usual 2HDM (see e.g.~\ccite{Branco:2011iw} for a review), the 2HDMa adds the following terms to the Lagrangian~\cite{Bauer:2017ota,LHCDarkMatterWorkingGroup:2018ufk},
\begin{align}
    \mathcal{L}_\text{2HDMa} ={}& \mathcal{L}_\text{2HDM} + \bar\chi (\slashed{\partial} + m_\chi) \chi - y_\chi a_0 \bar\chi i\gamma^5 \chi \nonumber\\
    & + \frac{1}{2}\partial_\mu a_0 \partial^\mu a_0 - \frac{1}{2}m_{a_0}^2 a_0^2 + i \kappa a_0 \Phi_1^\dagger \Phi_2 + \text{h.c.} + \ldots, 
\end{align}
where $a_0$ is the additional gauge-singlet pseudoscalar of mass $m_{a_0}$, $\chi$ is a Dirac fermion dark matter candidate of mass $m_\chi$, and $\Phi_{1,2}$ are the two 2HDM doublets. The dark sector Yukawa coupling is denoted as $y_\chi$. The portal coupling $\kappa$ and the pseudoscalar mass $m_{a_0}$ are parameters with mass dimension one. The ellipsis denotes additional quartic interactions which are not relevant for our discussion here. After electroweak symmetry breaking, the 2HDM pseudoscalar, which we denote by $A_0$, will mix with $a_0$ (with the mixing angle $\theta$) leading to the mass eigenstates $a$ and $A$.

This model has been used as a possible explanation for the gamma ray Galactic Centre excess observed by the Fermi-LAT space telescope~\cite{Fermi-LAT:2015sau,Izaguirre:2014vva,Ipek:2014gua} and for the baryon asymmetry of the Universe~\cite{Huber:2022ndk}. Moreover, this model is often used as a benchmark model for LHC dark-matter searches~\cite{LHCDarkMatterWorkingGroup:2018ufk}. As discussed in \ccite{Tunney:2017yfp}, the LHC signature
\begin{align}\label{eq:2HDM_process}
    p p \to b \bar b H \to b\bar b Z a\to b \bar b l^+l^- \chi\bar\chi
\end{align}
could be a promising way to test this explanation of the Galactic Centre excess, which so far cannot be probed by other constraints.

This signature can be straightforwardly mapped to our 2-vs-1 unbalanced simplified model topology: $H$ plays the role of the scalar resonance $\Phi$; $a$ is the neutral mediator $M$; and $\chi$ is the invisible particle $I$. This mapping allows us to easily reinterpret the model-independent expected limits derived above within the 2HDMa parameter space.

We calculate the cross section for the process in \cref{eq:2HDM_process} at the leading order using \texttt{MadGraph5\_aMC@NLO} (v2.6.5).  These cross section values are then compared to the expected limits presented in the lower left panel of \cref{fig:Limits_2D}. We also show projections for increased luminosities by simply rescaling the limits by the square root of the relative increase in luminosity.

\begin{figure}
    \centering
    \includegraphics[width=0.5\textwidth]{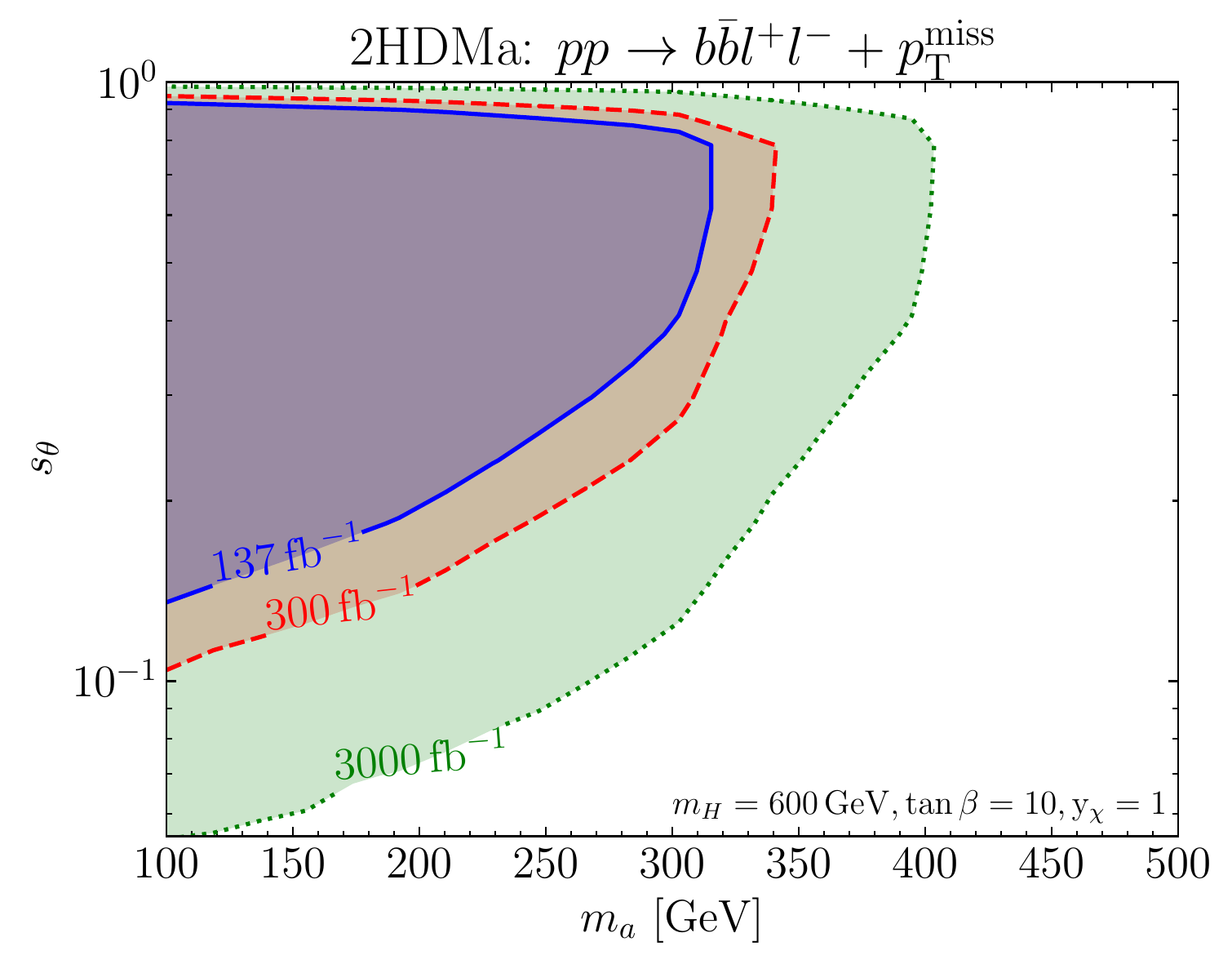}\hspace{-.4cm}
    \includegraphics[width=0.5\textwidth]{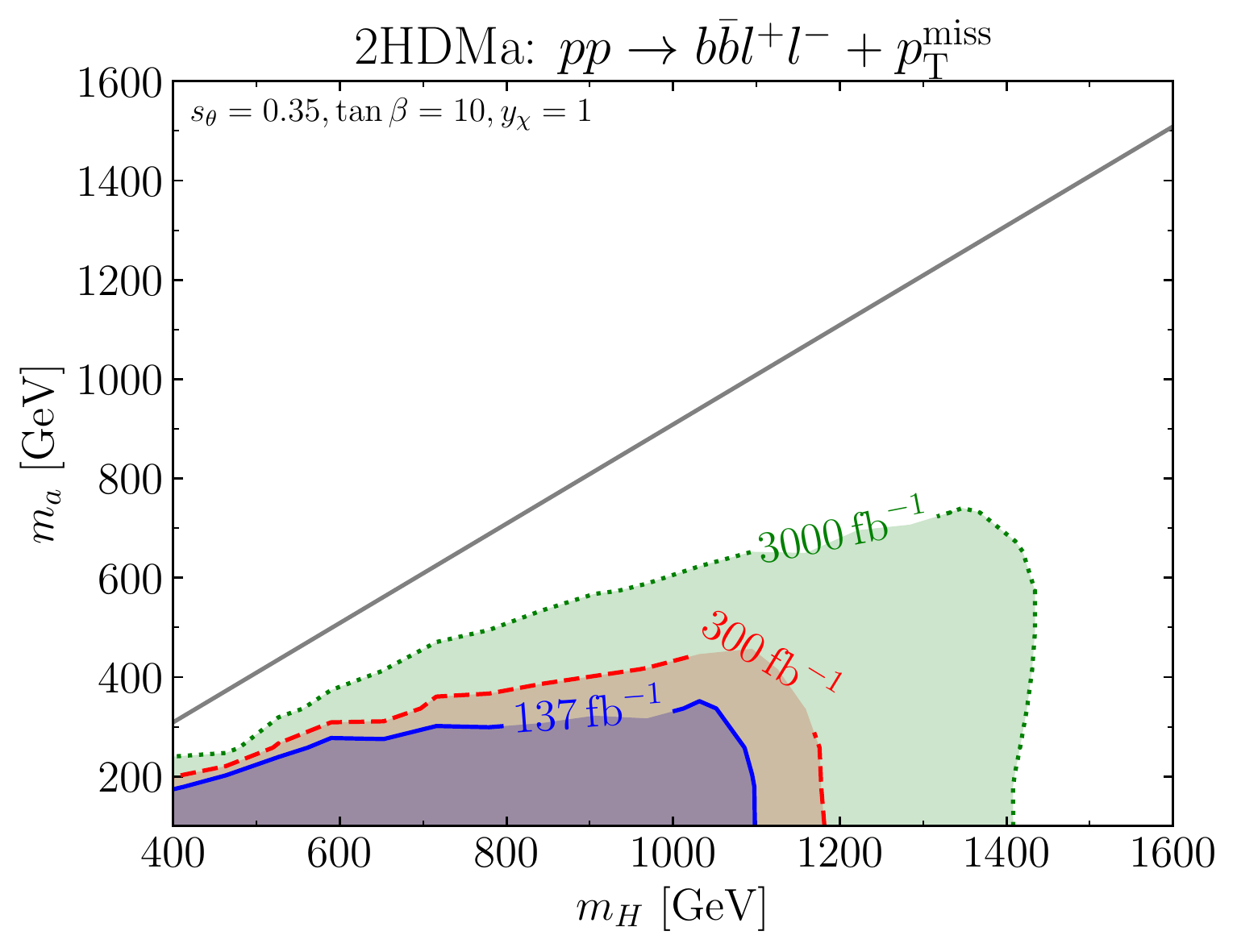}
    \includegraphics[width=0.5\textwidth]{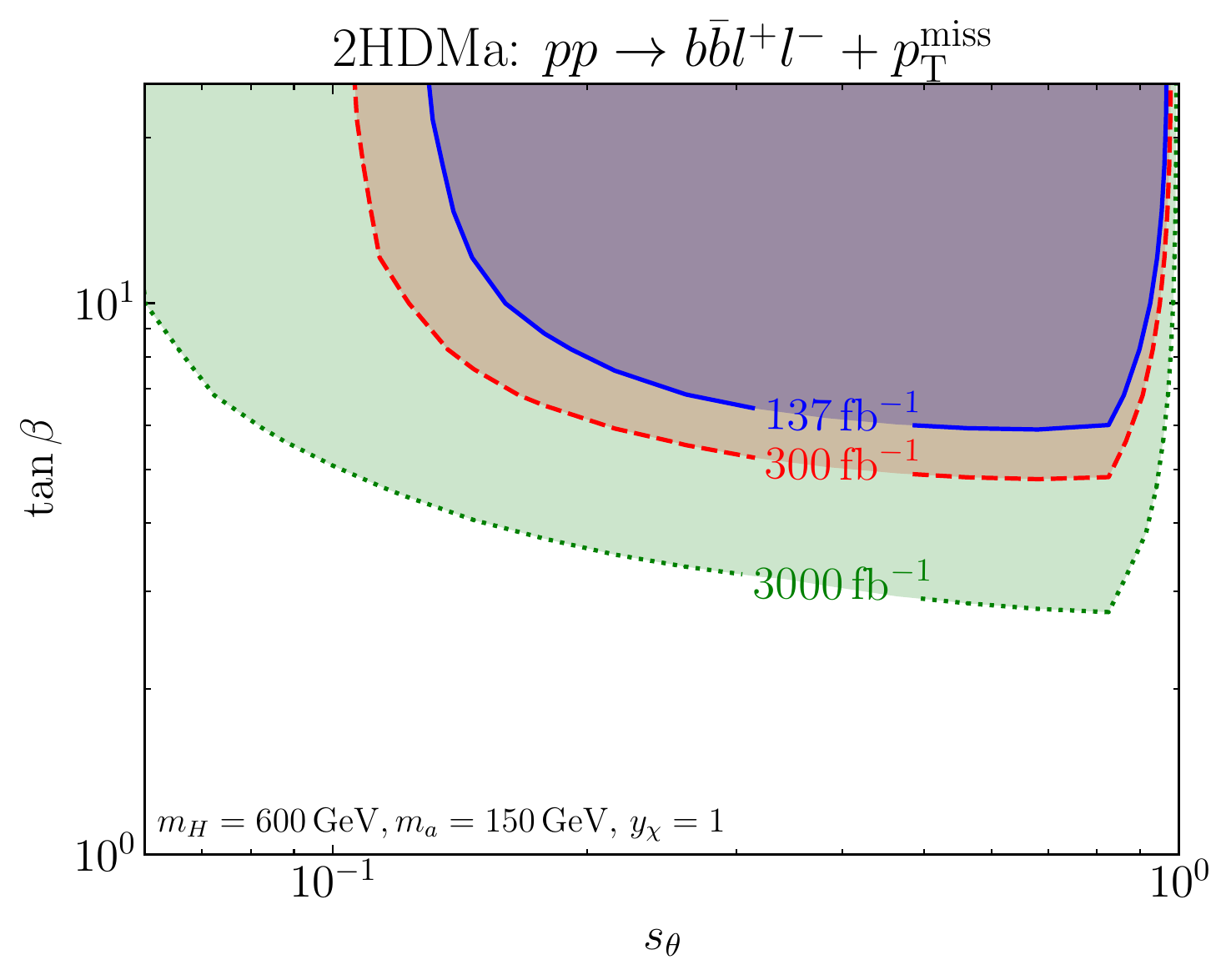}\hspace{-.4cm}
    \includegraphics[width=0.5\textwidth]{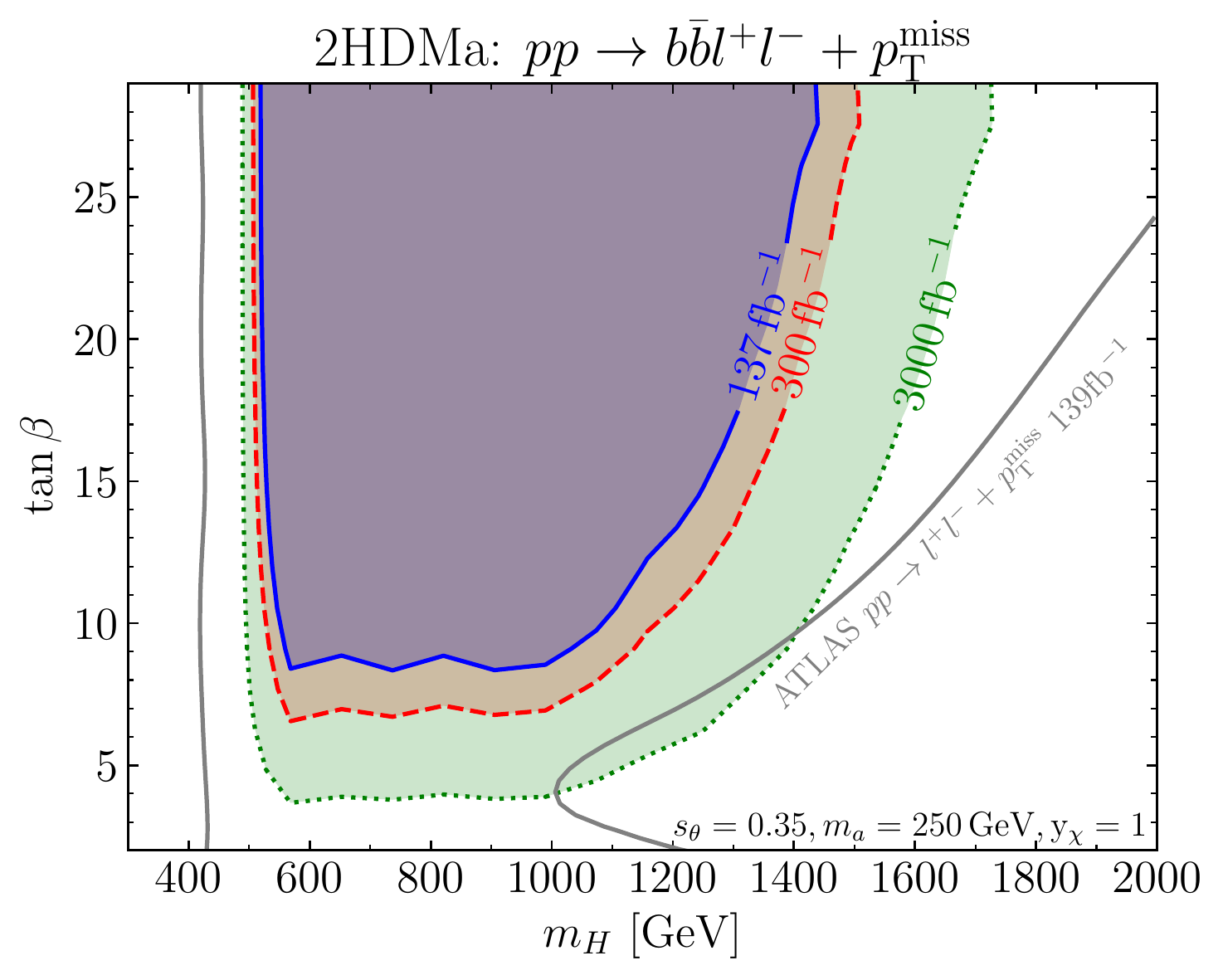}
    \caption{ 
    Projected expected limits on the 2HDMa parameter space in the $(m_a,s_\theta)$ parameter plane (upper left), in the $(m_H, m_a)$ parameter plane (upper right), in the $(s_\theta,\tan\beta)$ parameter plane (lower left), and in the $(m_H,\tan\beta)$ parameter plane (lower right). The limits are shown for three different luminosities: $137\invfb$ (blue), $300\invfb$ (red), and $3000\invfb$ (green). For the lower right parameter plane, we also show the limit presented in \ccite{ATLAS:2021gcn}.}
    \label{fig:2HDMa_lims}
\end{figure}

\cref{fig:2HDMa_lims} shows the resulting constraints on the 2HDMa parameter space in four different parameter planes. In the upper left panel, we show the $(m_a, s_\theta)$ parameter plane with $m_H = 600\gev$ and $\tan\beta = 10$. With a luminosity of $137\invfb$, pseudoscalar masses of up to $300\gev$ can be probed; $s_\theta$ can be probed down to values of $\sim 0.15$.\footnote{For $s_\theta\to 1$ the cross section for the process in \cref{eq:2HDM_process} goes to zero resulting in a rapid loss of sensitivity.} Increasing the luminosity enhances the sensitivity, especially in the low $s_\theta$ region. Our projected limit is in good agreement with the projection obtained in \ccite{Tunney:2017yfp} confirming the potential to probe models that provide explanations for the Galactic Centre excess.\footnote{Small differences can be explained by the inclusion of a $K$ factor of 1.4 for the bottom-associated Higgs production cross section and by a more simplified treatment of detector effects in \ccite{Tunney:2017yfp}. The selection cuts  in \ccite{Tunney:2017yfp} are, moreover, optimized for the unbalanced topology.}

The limits are shown in the $(m_H, m_a)$ parameter plane fixing $s_\theta=0.35$ and $\tan\beta=10$ in the upper right panel of \cref{fig:2HDMa_lims}. The gray line delimits the physical region for which $m_H \ge m_a + m_Z$. With a luminosity of $137\invfb$ the proposed search can probe scalar masses up to $\sim 1.1\tev$. Increasing the luminosity to $300\invfb$ ($3000\invfb$), masses up to $1.2\tev$ ($1.4\tev$) are tested.

In the 2HDMa, the bottom-associated $H$ production is enhanced for large $\tan\beta$. Correspondingly, stronger limits in the large $\tan\beta$ region are expected. This is confirmed in the lower left panel of \cref{fig:2HDMa_lims}, which shows the limits in the $(s_\theta, \tan\beta)$ parameter plane for $m_H = 600\gev$ and $m_a = 150\gev$. As expected from the dependence of the cross-section on $s_\theta$ and $\tan\beta$, the limits are strongest for $s_\theta \sim 0.9$ and large $\tan\beta$. With a luminosity of $137\invfb$, $\tan\beta$ values down to $\sim 6$ can be probed for $s_\theta \sim 0.9$. Increasing the luminosity to $300\invfb$ ($3000\invfb$), $\tan\beta$ values down to $\sim 5$ ($\sim 3$) are tested.

The dependence on $\tan\beta$ is further explored in the lower right panel of \cref{fig:2HDMa_lims}, which shows the limits in the $(m_H, \tan\beta)$ plane for $s_\theta = 0.35$ and $m_a = 250\gev$. While the proposed search is not able to probe the region of $\tan\beta \lesssim 3$ even with $\mathcal{L}=3000\invfb$, the interval $500\gev \lesssim m_H \lesssim 1600\gev$ can be probed for larger $\tan\beta$. For comparison, we also show the 2HDMa limit set by the search presented by the ATLAS Collaboration in \ccite{ATLAS:2021gcn} with a luminosity of $139\invfb$. We observe that this limit is significantly stronger than our projection for $\mathcal{L}=137\invfb$ and $\mathcal{L}=300\invfb$. Moreover, it also outperforms our projection for $\mathcal{L}=3000\invfb$ except for a parameter region with relatively small $\tan\beta$. This difference in sensitivity is mainly a consequence of the different final states considered in our analysis and in \ccite{ATLAS:2021gcn}. In contrast to our proposed search, \ccite{ATLAS:2021gcn} vetoes $b$-jets in the final state, searching for a $Z + \pTmiss$ final state. While the $b\bar b Z + \pTmiss$ final state, which our analysis targets, allows one to identify the production mode in case a signal is detected, the $Z + \pTmiss$ final state seems to have better sensitivity in comparison to the bottom-associated Higgs production in the case of the 2HDMa. Moreover, it should be noted that our simplified analysis relies on a simplified detector simulation, is not optimized for the unbalanced topology, and, moreover, does not use advanced multivariate analysis techniques.

%%%%%%%%%%%%%%%%%%%%%%%%%%%%%%%%%%%%%%%%%%%%%%%%%%%%%%%%%%%%%%%%%%%%%%%%%%%%%%
%%%%%%%%%%%%%%%%%%%%%%%%%%%%%%%%%%%%%%%%%%%%%%%%%%%%%%%%%%%%%%%%%%%%%%%%%%%%%%

\section{Conclusions}
\label{sec:conclusions}

While it is expected that the production of DM at colliders will result in an increased amount of $\pTmiss$ in the detectors, the exact way in which this additional source of missing momentum will manifest itself is of course not established. Most existing searches are focused on well-motivated but at the same time specific models with a determined unbalanced event topology, for which the missing transverse momentum recoils against the visible particles. This restriction reduces the chance for a discovery, since other BSM models may predict more balanced event topologies.

In this work, we have studied the capability of a potential new search in the context of a greater variety of theoretical scenarios based on a flexible simplified model framework, which covers two unbalanced and two balanced event topologies. Our detailed collider study focuses on bottom-quark associated neutral Higgs production with a $b\bar b Z + \pTmiss$ final state. The analysis strategy exploited is based on a selection more inclined towards unbalanced configurations between the 
$\pTmiss$ and the recoiling product. 

Our results indicate that in the presence of the $b$-associated production of a heavy scalar, the inclusion of jets in the forward region of the detectors plays a very important role in the performance of the analysis. This yields for the study that is presented here a comparable sensitivity to the one expected using only the standard central jets with $b$-tagging.

 Furthermore, the obtained limits show that the analysis proposed here is well suited for both unbalanced topologies, yielding almost exactly the same results in both cases, and thus demonstrating that the kinematics of the two unbalanced topologies cannot be distinguished with the proposed analysis (which on the other hand implies that the same search can be applied to a wider class of theoretical models than the ones addressed by a dedicated search within a specific model). We showed that the analysis has a significant sensitivity also for the two balanced topologies, although this sensitivity is somewhat reduced as compared to the unbalanced case. Such scenarios, where there is a marked difference in kinematics with respect to the unbalanced topologies, will probably further profit from a dedicated strategy addressing the challenges arising when using softer objects and a more relaxed event selection. In the other extreme, where the $Z$ bosons become highly boosted and the reconstruction of individual leptons starts to fail, complementary searches like $b\bar{b}+\text{DM}$ could be of vital importance. As an additional application of the model-independent results that we have achieved, we presented projected limits on the 2HDMa parameter space. We find the projected limits to be not as strong as existing searches that do not require the presence of $b$-jets in the final state. Those searches are, however, more optimized for the unbalanced topology and use advanced multivariate analysis techniques.

 In this paper we have demonstrated for the example of the considered search how results for BSM Higgs searches can be presented in the framework of simplified models, i.e.\ within a framework that up to now has been applied very successfully for direct searches for BSM particles and DM, but so far has not been used by the experimental collaborations for BSM Higgs searches. We encourage the experimental collaborations to make use of this framework for the presentation of their future experimental results.

Finally, we hope that the presented study serves as a starting point for future experimental efforts which are not restricted to only unbalanced event topologies. More sophisticated experimental analysis techniques, such as the inclusion of machine learning and advanced algorithms to discriminate against the main background processes, could certainly be an important asset when performing such analyses in a more realistic context.

%%%%%%%%%%%%%%%%%%%%%%%%%%%%%%%%%%%%%%%%%%%%%%%%%%%%%%%%%%%%%%%%%%%%%%%%%%%%%%
%%%%%%%%%%%%%%%%%%%%%%%%%%%%%%%%%%%%%%%%%%%%%%%%%%%%%%%%%%%%%%%%%%%%%%%%%%%%%%

\section*{Acknowledgements}
\sloppy{
We thank Jose Miguel No and Katherina Behr for useful discussions. D.P.A., A.G., V.M.L., C.S., and G.W.\ acknowledge support by the Deutsche Forschungsgemeinschaft (DFG, German Research Foundation) under Germany's Excellence Strategy --- EXC 2121 “Quantum Universe” --- 390833306. H.B.\ acknowledges support from the Alexander von Humboldt foundation. V.M.L.\ acknowledges the financial support by Ministerio de Universidades and “European Union-NextGenerationEU/PRTR” under the grant María Zambrano ZA2021-081, and the Spanish grants PID2020-113775GB-I00 (AEI/10.13039/501100011033) and CIPROM/2021/054 (Generalitat Valenciana). This work has been partially funded by the Deutsche Forschungsgemeinschaft (DFG, German Research Foundation) --- 491245950.
}

%%%%%%%%%%%%%%%%%%%%%%%%%%%%%%%%%%%%%%%%%%%%%%%%%%%%%%%%%%%%%%%%%%%%%%%%%%%%%%
%%%%%%%%%%%%%%%%%%%%%%%%%%%%%%%%%%%%%%%%%%%%%%%%%%%%%%%%%%%%%%%%%%%%%%%%%%%%%%
%%%%%%%%%%%%%%%%%%%%%%%%%%%%%%%%%%%%%%%%%%%%%%%%%%%%%%%%%%%%%%%%%%%%%%%%%%%%%%

\clearpage
\printbibliography

@article{Arcadi:2019lka,
    author = "Arcadi, Giorgio and Djouadi, Abdelhak and Raidal, Martti",
    title = "{Dark Matter through the Higgs portal}",
    eprint = "1903.03616",
    archivePrefix = "arXiv",
    primaryClass = "hep-ph",
    reportNumber = "LAPTH-010/19",
    doi = "10.1016/j.physrep.2019.11.003",
    journal = "Phys. Rept.",
    volume = "842",
    pages = "1--180",
    year = "2020"
}

@article{Curtin:2013fra,
    author = "Curtin, David and others",
    title = "{Exotic decays of the 125 GeV Higgs boson}",
    eprint = "1312.4992",
    archivePrefix = "arXiv",
    primaryClass = "hep-ph",
    reportNumber = "YITP-13-47, PITT-PACC-1314",
    doi = "10.1103/PhysRevD.90.075004",
    journal = "Phys. Rev. D",
    volume = "90",
    number = "7",
    pages = "075004",
    year = "2014"
}

@article{Duerr:2016tmh,
    author = "Duerr, Michael and Kahlhoefer, Felix and Schmidt-Hoberg, Kai and Schwetz, Thomas and Vogl, Stefan",
    title = "{How to save the WIMP: global analysis of a dark matter model with two s-channel mediators}",
    eprint = "1606.07609",
    archivePrefix = "arXiv",
    primaryClass = "hep-ph",
    reportNumber = "DESY-16-113",
    doi = "10.1007/JHEP09(2016)042",
    journal = "JHEP",
    volume = "09",
    pages = "042",
    year = "2016"
}

@article{Feng:2010gw,
    author = "Feng, Jonathan L.",
    title = "{Dark Matter Candidates from Particle Physics and Methods of Detection}",
    eprint = "1003.0904",
    archivePrefix = "arXiv",
    primaryClass = "astro-ph.CO",
    reportNumber = "UCI-TR-2009-13",
    doi = "10.1146/annurev-astro-082708-101659",
    journal = "Ann. Rev. Astron. Astrophys.",
    volume = "48",
    pages = "495--545",
    year = "2010"
}

@article{Bertone:2004pz,
    author = "Bertone, Gianfranco and Hooper, Dan and Silk, Joseph",
    title = "{Particle dark matter: Evidence, candidates and constraints}",
    eprint = "hep-ph/0404175",
    archivePrefix = "arXiv",
    reportNumber = "FERMILAB-PUB-04-047-A",
    doi = "10.1016/j.physrep.2004.08.031",
    journal = "Phys. Rept.",
    volume = "405",
    pages = "279--390",
    year = "2005"
}

@article{ATLAS:2021kxv,
    author = "Aad, Georges and others",
    collaboration = "ATLAS",
    title = "{Search for new phenomena in events with an energetic jet and missing transverse momentum in $pp$ collisions at $\sqrt {s}$ =13  TeV with the ATLAS detector}",
    eprint = "2102.10874",
    archivePrefix = "arXiv",
    primaryClass = "hep-ex",
    reportNumber = "CERN-EP-2020-238",
    doi = "10.1103/PhysRevD.103.112006",
    journal = "Phys. Rev. D",
    volume = "103",
    number = "11",
    pages = "112006",
    year = "2021"
}

@article{CMS:2021far,
    author = "Tumasyan, Armen and others",
    collaboration = "CMS",
    title = "{Search for new particles in events with energetic jets and large missing transverse momentum in proton-proton collisions at $ \sqrt{s} $ = 13 TeV}",
    eprint = "2107.13021",
    archivePrefix = "arXiv",
    primaryClass = "hep-ex",
    reportNumber = "CMS-EXO-20-004, CERN-EP-2021-136",
    doi = "10.1007/JHEP11(2021)153",
    journal = "JHEP",
    volume = "11",
    pages = "153",
    year = "2021"
}

@article{ATLAS:2020uiq,
    author = "Aad, Georges and others",
    collaboration = "ATLAS",
    title = "{Search for dark matter in association with an energetic photon in $pp$ collisions at $\sqrt{s}$ = 13 TeV with the ATLAS detector}",
    eprint = "2011.05259",
    archivePrefix = "arXiv",
    primaryClass = "hep-ex",
    reportNumber = "CERN-EP-2020-178",
    doi = "10.1007/JHEP02(2021)226",
    journal = "JHEP",
    volume = "02",
    pages = "226",
    year = "2021"
}

@article{CMS:2018ffd,
    author = "Sirunyan, Albert M and others",
    collaboration = "CMS",
    title = "{Search for new physics in final states with a single photon and missing transverse momentum in proton-proton collisions at $\sqrt{s} =$ 13 TeV}",
    eprint = "1810.00196",
    archivePrefix = "arXiv",
    primaryClass = "hep-ex",
    reportNumber = "CMS-EXO-16-053, CERN-EP-2018-248",
    doi = "10.1007/JHEP02(2019)074",
    journal = "JHEP",
    volume = "02",
    pages = "074",
    year = "2019"
}

@article{ATLAS:2021gcn,
    author = "Aad, Georges and others",
    collaboration = "ATLAS",
    title = "{Search for associated production of a $Z$ boson with an invisibly decaying Higgs boson or dark matter candidates at $\sqrt s$ =13 TeV with the ATLAS detector}",
    eprint = "2111.08372",
    archivePrefix = "arXiv",
    primaryClass = "hep-ex",
    reportNumber = "CERN-EP-2021-204",
    doi = "10.1016/j.physletb.2022.137066",
    journal = "Phys. Lett. B",
    volume = "829",
    pages = "137066",
    year = "2022"
}

@article{CMS:2020ulv,
    author = "Sirunyan, Albert M and others",
    collaboration = "CMS",
    title = "{Search for dark matter produced in association with a leptonically decaying Z boson in proton-proton collisions at $\sqrt{s} =$ 13 TeV}",
    eprint = "2008.04735",
    archivePrefix = "arXiv",
    primaryClass = "hep-ex",
    reportNumber = "CMS-EXO-19-003, CERN-EP-2020-136",
    doi = "10.1140/epjc/s10052-020-08739-5",
    journal = "Eur. Phys. J. C",
    volume = "81",
    number = "1",
    pages = "13",
    year = "2021",
    note = "[Erratum: Eur.Phys.J.C 81, 333 (2021)]"
}

@article{ATLAS:2021shl,
    author = "Aad, Georges and others",
    collaboration = "ATLAS",
    title = "{Search for dark matter produced in association with a Standard Model Higgs boson decaying into b-quarks using the full Run 2 dataset from the ATLAS detector}",
    eprint = "2108.13391",
    archivePrefix = "arXiv",
    primaryClass = "hep-ex",
    reportNumber = "CERN-EP-2021-074",
    doi = "10.1007/JHEP11(2021)209",
    journal = "JHEP",
    volume = "11",
    pages = "209",
    year = "2021"
}

@article{ATLAS:2021jbf,
    author = "Aad, Georges and others",
    collaboration = "ATLAS",
    title = "{Search for dark matter in events with missing transverse momentum and a Higgs boson decaying into two photons in pp collisions at $ \sqrt{s} $ = 13 TeV with the ATLAS detector}",
    eprint = "2104.13240",
    archivePrefix = "arXiv",
    primaryClass = "hep-ex",
    reportNumber = "CERN-EP-2021-012",
    doi = "10.1007/JHEP10(2021)013",
    journal = "JHEP",
    volume = "10",
    pages = "013",
    year = "2021"
}

@article{CMS:2019ykj,
    author = "Sirunyan, Albert M and others",
    collaboration = "CMS",
    title = "{Search for dark matter particles produced in association with a Higgs boson in proton-proton collisions at $ \sqrt{\mathrm{s}} $ = 13 TeV}",
    eprint = "1908.01713",
    archivePrefix = "arXiv",
    primaryClass = "hep-ex",
    reportNumber = "CMS-EXO-18-011, CERN-EP-2019-141",
    doi = "10.1007/JHEP03(2020)025",
    journal = "JHEP",
    volume = "03",
    pages = "025",
    year = "2020"
}

@article{CMS:2018zjv,
    author = "Sirunyan, Albert M and others",
    collaboration = "CMS",
    title = "{Search for dark matter produced in association with a Higgs boson decaying to a pair of bottom quarks in proton\textendash{}proton collisions at $\sqrt{s}=13\,\text {Te}\text {V} $}",
    eprint = "1811.06562",
    archivePrefix = "arXiv",
    primaryClass = "hep-ex",
    reportNumber = "CMS-EXO-16-050, CERN-EP-2018-287",
    doi = "10.1140/epjc/s10052-019-6730-7",
    journal = "Eur. Phys. J. C",
    volume = "79",
    number = "3",
    pages = "280",
    year = "2019"
}

@article{CMS:2018nlv,
    author = "Sirunyan, Albert M. and others",
    collaboration = "CMS",
    title = "{Search for dark matter produced in association with a Higgs boson decaying to $\gamma\gamma$ or $\tau^+\tau^-$ at $\sqrt{s} =$ 13 TeV}",
    eprint = "1806.04771",
    archivePrefix = "arXiv",
    primaryClass = "hep-ex",
    reportNumber = "CMS-EXO-16-055, CERN-EP-2018-129",
    doi = "10.1007/JHEP09(2018)046",
    journal = "JHEP",
    volume = "09",
    pages = "046",
    year = "2018"
}

@article{ATLAS:2022bzt,
    collaboration = "ATLAS",
    title = "{Search for dark matter produced in association with a dark Higgs boson decaying into $W^{+}W^{-}$ in the one-lepton final state at $\sqrt{s}$=13 TeV using 139 fb$^{-1}$ of $pp$ collisions recorded with the ATLAS detector}",
    eprint = "2211.07175",
    archivePrefix = "arXiv",
    primaryClass = "hep-ex",
    reportNumber = "CERN-EP-2022-147",
    month = "11",
    year = "2022"
}

@article{ATLAS:2020fgc,
    author = "Aad, Georges and others",
    collaboration = "ATLAS",
    title = "{Search for Dark Matter Produced in Association with a Dark Higgs Boson Decaying into $W^\pm W^\mp$ or $ZZ$ in Fully Hadronic Final States from $\sqrt{s}=13$ TeV pp Collisions Recorded with the ATLAS Detector}",
    eprint = "2010.06548",
    archivePrefix = "arXiv",
    primaryClass = "hep-ex",
    reportNumber = "CERN-EP-2020-172",
    doi = "10.1103/PhysRevLett.126.121802",
    journal = "Phys. Rev. Lett.",
    volume = "126",
    number = "12",
    pages = "121802",
    year = "2021"
}

@article{ATLAS:2022znu,
    collaboration = "ATLAS",
    title = "{Search for dark matter produced in association with a single top quark and an energetic $W$ boson in $\sqrt{s}=$ 13 TeV $pp$ collisions with the ATLAS detector}",
    eprint = "2211.13138",
    archivePrefix = "arXiv",
    primaryClass = "hep-ex",
    reportNumber = "CERN-EP-2022-146",
    month = "11",
    year = "2022"
}

@article{ATLAS:2020yzc,
    author = "Aad, Georges and others",
    collaboration = "ATLAS",
    title = "{Search for dark matter produced in association with a single top quark in $\sqrt{s}=13$~TeV $pp$ collisions with the ATLAS detector}",
    eprint = "2011.09308",
    archivePrefix = "arXiv",
    primaryClass = "hep-ex",
    reportNumber = "CERN-EP-2020-184",
    doi = "10.1140/epjc/s10052-021-09566-y",
    journal = "Eur. Phys. J. C",
    volume = "81",
    pages = "860",
    year = "2021"
}

@article{CMS:2019zzl,
    author = "Sirunyan, Albert M and others",
    collaboration = "CMS",
    title = "{Search for dark matter produced in association with a single top quark or a top quark pair in proton-proton collisions at $ \sqrt{s}=13 $ TeV}",
    eprint = "1901.01553",
    archivePrefix = "arXiv",
    primaryClass = "hep-ex",
    reportNumber = "CMS-EXO-18-010, CERN-EP-2018-311",
    doi = "10.1007/JHEP03(2019)141",
    journal = "JHEP",
    volume = "03",
    pages = "141",
    year = "2019"
}

@article{CMS:2018ysw,
    author = "Sirunyan, Albert M. and others",
    collaboration = "CMS",
    title = "{Search for dark matter particles produced in association with a top quark pair at $\sqrt{s} =$ 13 TeV}",
    eprint = "1807.06522",
    archivePrefix = "arXiv",
    primaryClass = "hep-ex",
    reportNumber = "CMS-EXO-16-049, CERN-EP-2018-183",
    doi = "10.1103/PhysRevLett.122.011803",
    journal = "Phys. Rev. Lett.",
    volume = "122",
    number = "1",
    pages = "011803",
    year = "2019"
}

@article{Abdallah:2015ter,
    author = "Abdallah, Jalal and others",
    title = "{Simplified Models for Dark Matter Searches at the LHC}",
    eprint = "1506.03116",
    archivePrefix = "arXiv",
    primaryClass = "hep-ph",
    reportNumber = "FERMILAB-PUB-15-283-CD, CERN-PH-TH-2015-139",
    doi = "10.1016/j.dark.2015.08.001",
    journal = "Phys. Dark Univ.",
    volume = "9-10",
    pages = "8--23",
    year = "2015"
}

@article{Abercrombie:2015wmb,
    author = "Abercrombie, Daniel and others",
    editor = "Boveia, Antonio and Doglioni, Caterina and Lowette, Steven and Malik, Sarah and Mrenna, Stephen",
    title = "{Dark Matter benchmark models for early LHC Run-2 Searches: Report of the ATLAS/CMS Dark Matter Forum}",
    eprint = "1507.00966",
    archivePrefix = "arXiv",
    primaryClass = "hep-ex",
    reportNumber = "FERMILAB-PUB-15-282-CD",
    doi = "10.1016/j.dark.2019.100371",
    journal = "Phys. Dark Univ.",
    volume = "27",
    pages = "100371",
    year = "2020"
}

@article{LHCDarkMatterWorkingGroup:2018ufk,
    author = "Abe, Tomohiro and others",
    collaboration = "LHC Dark Matter Working Group",
    title = "{LHC Dark Matter Working Group: Next-generation spin-0 dark matter models}",
    eprint = "1810.09420",
    archivePrefix = "arXiv",
    primaryClass = "hep-ex",
    reportNumber = "CERN-LPCC-2018-02",
    doi = "10.1016/j.dark.2019.100351",
    journal = "Phys. Dark Univ.",
    volume = "27",
    pages = "100351",
    year = "2020"
}

@article{Buckley:2014ana,
    author = {Buckley, Andy and Ferrando, James and Lloyd, Stephen and Nordstr\"om, Karl and Page, Ben and R\"ufenacht, Martin and Sch\"onherr, Marek and Watt, Graeme},
    title = "{LHAPDF6: parton density access in the LHC precision era}",
    eprint = "1412.7420",
    archivePrefix = "arXiv",
    primaryClass = "hep-ph",
    reportNumber = "GLAS-PPE-2014-05, MCNET-14-29, IPPP-14-111, DCPT-14-222",
    doi = "10.1140/epjc/s10052-015-3318-8",
    journal = "Eur. Phys. J. C",
    volume = "75",
    pages = "132",
    year = "2015"
}

@article{Andersen:2014efa,
    author = "Andersen, J. R. and others",
    title = "{Les Houches 2013: Physics at TeV Colliders: Standard Model Working Group Report}",
    eprint = "1405.1067",
    archivePrefix = "arXiv",
    primaryClass = "hep-ph",
    month = "5",
    year = "2014"
}

@article{Bahl:2021xmy,
    author = "Bahl, Henning and Lozano, Victor Martin and Weiglein, Georg",
    title = "{Simplified models for resonant neutral scalar production with missing transverse energy final states}",
    eprint = "2112.12656",
    archivePrefix = "arXiv",
    primaryClass = "hep-ph",
    reportNumber = "DESY-21-227, EFI-12-10",
    doi = "10.1007/JHEP11(2022)042",
    journal = "JHEP",
    volume = "11",
    pages = "042",
    year = "2022"
}

@misc{Run2:Luminosity,
  collaboration = "CMS",
  title        = "Luminosity Public Results",
  url          = "https://twiki.cern.ch/twiki/bin/view/CMSPublic/LumiPublicResults",
  month        = "03",
  year         = "2022"
}

@article{CMS:2019jhq,
    collaboration = "CMS",
    title = "{CMS luminosity measurement for the 2018 data-taking period at $\sqrt{s} = 13~\mathrm{TeV}$}",
    reportNumber = "CMS-PAS-LUM-18-002",
    year = "2019"
}

@article{Czakon:2013goa,
    author = "Czakon, Micha\l{} and Fiedler, Paul and Mitov, Alexander",
    title = "{Total Top-Quark Pair-Production Cross Section at Hadron Colliders
Through $O(\alpha^4_S)$}",
    eprint = "1303.6254",
    archivePrefix = "arXiv",
    primaryClass = "hep-ph",
    reportNumber = "CERN-PH-TH-2013-056, TTK-13-08",
    doi = "10.1103/PhysRevLett.110.252004",
    journal = "Phys. Rev. Lett.",
    volume = "110",
    pages = "252004",
    year = "2013"
}

@inproceedings{Kidonakis:2013zqa,
    author = "Kidonakis, Nikolaos",
    title = "{Top Quark Production}",
    booktitle = "{Helmholtz International Summer School on Physics of Heavy Quarks and Hadrons}",
    eprint = "1311.0283",
    archivePrefix = "arXiv",
    primaryClass = "hep-ph",
    reportNumber = "KSU-HEP-110113",
    doi = "10.3204/DESY-PROC-2013-03/Kidonakis",
    pages = "139--168",
    year = "2014"
}

@article{Li:2012wna,
    author = "Li, Ye and Petriello, Frank",
    title = "{Combining QCD and electroweak corrections to dilepton production in FEWZ}",
    eprint = "1208.5967",
    archivePrefix = "arXiv",
    primaryClass = "hep-ph",
    reportNumber = "ANL-HEP-PR-12-68",
    doi = "10.1103/PhysRevD.86.094034",
    journal = "Phys. Rev. D",
    volume = "86",
    pages = "094034",
    year = "2012"
}

@misc{RooStats:Twiki,
  collaboration = "ROOT team and LHC experiments",
  title         = "The RooStats framework",
  url           = "https://twiki.cern.ch/twiki/bin/view/RooStats/WebHome",
  month         = "05",
  year          = "2022"
}

@techreport{Cranmer:1456844,
      author        = "Cranmer, Kyle and Lewis, George and Moneta, Lorenzo and
                       Shibata, Akira and Verkerke, Wouter",
      title         = "{HistFactory: A tool for creating statistical models for
                       use with RooFit and RooStats}",
      institution   = "New York U.",
      collaboration = "ROOT Collaboration",
      address       = "New York",
      reportNumber  = "CERN-OPEN-2012-016",
      month         = "1",
      year          = "2012",
      url           = "https://cds.cern.ch/record/1456844",
}

@article{Antcheva:2009zz,
    author = "Antcheva, I. and others",
    title = "{ROOT: A C++ framework for petabyte data storage, statistical analysis and visualization}",
    eprint = "1508.07749",
    archivePrefix = "arXiv",
    primaryClass = "physics.data-an",
    reportNumber = "FERMILAB-PUB-09-661-CD",
    doi = "10.1016/j.cpc.2009.08.005",
    journal = "Comput. Phys. Commun.",
    volume = "180",
    pages = "2499--2512",
    year = "2009"
}

@article{Cowan:2010js,
    author = "Cowan, Glen and Cranmer, Kyle and Gross, Eilam and Vitells, Ofer",
    title = "{Asymptotic formulae for likelihood-based tests of new physics}",
    eprint = "1007.1727",
    archivePrefix = "arXiv",
    primaryClass = "physics.data-an",
    doi = "10.1140/epjc/s10052-011-1554-0",
    journal = "Eur. Phys. J. C",
    volume = "71",
    pages = "1554",
    year = "2011",
    note = "[Erratum: Eur.Phys.J.C 73, 2501 (2013)]"
}

@article{Read:451614,
      author        = "Read, A L",
      title         = "{Modified frequentist analysis of search results (the
                       $CL_{s}$ method)}",
      reportNumber  = "CERN-OPEN-2000-205",
      year          = "2000",
      url           = "http://cds.cern.ch/record/451614",
      doi           = "10.5170/CERN-2000-005.81",
}

@article{Branco:2011iw,
    author = "Branco, G. C. and Ferreira, P. M. and Lavoura, L. and Rebelo, M. N. and Sher, Marc and Silva, Joao P.",
    title = "{Theory and phenomenology of two-Higgs-doublet models}",
    eprint = "1106.0034",
    archivePrefix = "arXiv",
    primaryClass = "hep-ph",
    doi = "10.1016/j.physrep.2012.02.002",
    journal = "Phys. Rept.",
    volume = "516",
    pages = "1--102",
    year = "2012"
}

@article{Tunney:2017yfp,
    author = "Tunney, Patrick and No, Jose Miguel and Fairbairn, Malcolm",
    title = "{Probing the pseudoscalar portal to dark matter via $\bar bbZ(\to\ell\ell)+ \slashed{E}_T$ : From the LHC to the Galactic Center excess}",
    eprint = "1705.09670",
    archivePrefix = "arXiv",
    primaryClass = "hep-ph",
    doi = "10.1103/PhysRevD.96.095020",
    journal = "Phys. Rev. D",
    volume = "96",
    number = "9",
    pages = "095020",
    year = "2017"
}

@article{Fermi-LAT:2015sau,
    author = "Ajello, M. and others",
    collaboration = "Fermi-LAT",
    title = "{Fermi-LAT Observations of High-Energy $\gamma$-Ray Emission Toward the Galactic Center}",
    eprint = "1511.02938",
    archivePrefix = "arXiv",
    primaryClass = "astro-ph.HE",
    doi = "10.3847/0004-637X/819/1/44",
    journal = "Astrophys. J.",
    volume = "819",
    number = "1",
    pages = "44",
    year = "2016"
}

@article{Izaguirre:2014vva,
    author = "Izaguirre, Eder and Krnjaic, Gordan and Shuve, Brian",
    title = "{The Galactic Center Excess from the Bottom Up}",
    eprint = "1404.2018",
    archivePrefix = "arXiv",
    primaryClass = "hep-ph",
    doi = "10.1103/PhysRevD.90.055002",
    journal = "Phys. Rev. D",
    volume = "90",
    number = "5",
    pages = "055002",
    year = "2014"
}

@article{Ipek:2014gua,
    author = "Ipek, Seyda and McKeen, David and Nelson, Ann E.",
    title = "{A Renormalizable Model for the Galactic Center Gamma Ray Excess from Dark Matter Annihilation}",
    eprint = "1404.3716",
    archivePrefix = "arXiv",
    primaryClass = "hep-ph",
    doi = "10.1103/PhysRevD.90.055021",
    journal = "Phys. Rev. D",
    volume = "90",
    number = "5",
    pages = "055021",
    year = "2014"
}

@article{Bauer:2017ota,
    author = "Bauer, Martin and Haisch, Ulrich and Kahlhoefer, Felix",
    title = "{Simplified dark matter models with two Higgs doublets: I. Pseudoscalar mediators}",
    eprint = "1701.07427",
    archivePrefix = "arXiv",
    primaryClass = "hep-ph",
    reportNumber = "CERN-TH-2017-011, DESY-17-010",
    doi = "10.1007/JHEP05(2017)138",
    journal = "JHEP",
    volume = "05",
    pages = "138",
    year = "2017"
}

@article{ATLAS:2012yve,
    author = "Aad, Georges and others",
    collaboration = "ATLAS",
    title = "{Observation of a new particle in the search for the Standard Model Higgs boson with the ATLAS detector at the LHC}",
    eprint = "1207.7214",
    archivePrefix = "arXiv",
    primaryClass = "hep-ex",
    reportNumber = "CERN-PH-EP-2012-218",
    doi = "10.1016/j.physletb.2012.08.020",
    journal = "Phys. Lett. B",
    volume = "716",
    pages = "1--29",
    year = "2012"
}

@article{CMS:2012qbp,
    author = "Chatrchyan, Serguei and others",
    collaboration = "CMS",
    title = "{Observation of a New Boson at a Mass of 125 GeV with the CMS Experiment at the LHC}",
    eprint = "1207.7235",
    archivePrefix = "arXiv",
    primaryClass = "hep-ex",
    reportNumber = "CMS-HIG-12-028, CERN-PH-EP-2012-220",
    doi = "10.1016/j.physletb.2012.08.021",
    journal = "Phys. Lett. B",
    volume = "716",
    pages = "30--61",
    year = "2012"
}

@article{CMS:2008xjf,
    author = "Chatrchyan, S. and others",
    collaboration = "CMS",
    title = "{The CMS Experiment at the CERN LHC}",
    doi = "10.1088/1748-0221/3/08/S08004",
    journal = "JINST",
    volume = "3",
    pages = "S08004",
    year = "2008"
}

@article{Alwall:2014hca,
    author = "Alwall, J. and Frederix, R. and Frixione, S. and Hirschi, V. and Maltoni, F. and Mattelaer, O. and Shao, H. -S. and Stelzer, T. and Torrielli, P. and Zaro, M.",
    title = "{The automated computation of tree-level and next-to-leading order differential cross sections, and their matching to parton shower simulations}",
    eprint = "1405.0301",
    archivePrefix = "arXiv",
    primaryClass = "hep-ph",
    reportNumber = "CERN-PH-TH-2014-064, CP3-14-18, LPN14-066, MCNET-14-09, ZU-TH-14-14",
    doi = "10.1007/JHEP07(2014)079",
    journal = "JHEP",
    volume = "07",
    pages = "079",
    year = "2014"
}

@article{Frederix:2012ps,
    author = "Frederix, Rikkert and Frixione, Stefano",
    title = "{Merging meets matching in MC@NLO}",
    eprint = "1209.6215",
    archivePrefix = "arXiv",
    primaryClass = "hep-ph",
    reportNumber = "CERN-PH-TH-2012-247, ZU-TH-21-12",
    doi = "10.1007/JHEP12(2012)061",
    journal = "JHEP",
    volume = "12",
    pages = "061",
    year = "2012"
}

@article{Sjostrand:2014zea,
    author = {Sj\"ostrand, Torbj\"orn and Ask, Stefan and Christiansen, Jesper R. and Corke, Richard and Desai, Nishita and Ilten, Philip and Mrenna, Stephen and Prestel, Stefan and Rasmussen, Christine O. and Skands, Peter Z.},
    title = "{An introduction to PYTHIA 8.2}",
    eprint = "1410.3012",
    archivePrefix = "arXiv",
    primaryClass = "hep-ph",
    reportNumber = "LU-TP-14-36, MCNET-14-22, CERN-PH-TH-2014-190, FERMILAB-PUB-14-316-CD, DESY-14-178, SLAC-PUB-16122",
    doi = "10.1016/j.cpc.2015.01.024",
    journal = "Comput. Phys. Commun.",
    volume = "191",
    pages = "159--177",
    year = "2015"
}

@article{NNPDF:2017mvq,
    author = "Ball, Richard D. and others",
    collaboration = "NNPDF",
    title = "{Parton distributions from high-precision collider data}",
    eprint = "1706.00428",
    archivePrefix = "arXiv",
    primaryClass = "hep-ph",
    reportNumber = "EDINBURGH-2017-08, NIKHEF-2017-006, OUTP-17-04P, TIF-UNIMI-2017-3, CAVENDISH-HEP-17-06, CERN-TH-2017-077, Edinburgh 2017/08,
  Nikhef/2017-006, OUTP-17-04P,TIF-UNIMI-2017-3",
    doi = "10.1140/epjc/s10052-017-5199-5",
    journal = "Eur. Phys. J. C",
    volume = "77",
    number = "10",
    pages = "663",
    year = "2017"
}

@article{Alioli:2010xd,
    author = "Alioli, Simone and Nason, Paolo and Oleari, Carlo and Re, Emanuele",
    title = "{A general framework for implementing NLO calculations in shower Monte Carlo programs: the POWHEG BOX}",
    eprint = "1002.2581",
    archivePrefix = "arXiv",
    primaryClass = "hep-ph",
    reportNumber = "DESY-10-018, SFB-CPP-10-22, IPPP-10-11, DCPT-10-22",
    doi = "10.1007/JHEP06(2010)043",
    journal = "JHEP",
    volume = "06",
    pages = "043",
    year = "2010"
}

@article{Frixione:2007nw,
    author = "Frixione, Stefano and Nason, Paolo and Ridolfi, Giovanni",
    title = "{A Positive-weight next-to-leading-order Monte Carlo for heavy flavour hadroproduction}",
    eprint = "0707.3088",
    archivePrefix = "arXiv",
    primaryClass = "hep-ph",
    reportNumber = "BICOCCA-FT-07-12, GEF-TH-19-2007",
    doi = "10.1088/1126-6708/2007/09/126",
    journal = "JHEP",
    volume = "09",
    pages = "126",
    year = "2007"
}

@article{deFavereau:2013fsa,
    author = "de Favereau, J. and Delaere, C. and Demin, P. and Giammanco, A. and Lema\^\i{}tre, V. and Mertens, A. and Selvaggi, M.",
    collaboration = "DELPHES 3",
    title = "{DELPHES 3, A modular framework for fast simulation of a generic collider experiment}",
    eprint = "1307.6346",
    archivePrefix = "arXiv",
    primaryClass = "hep-ex",
    doi = "10.1007/JHEP02(2014)057",
    journal = "JHEP",
    volume = "02",
    pages = "057",
    year = "2014"
}

@article{Conte:2012fm,
    author = "Conte, Eric and Fuks, Benjamin and Serret, Guillaume",
    title = "{MadAnalysis 5, A User-Friendly Framework for Collider Phenomenology}",
    eprint = "1206.1599",
    archivePrefix = "arXiv",
    primaryClass = "hep-ph",
    reportNumber = "IPHC-PHENO-06",
    doi = "10.1016/j.cpc.2012.09.009",
    journal = "Comput. Phys. Commun.",
    volume = "184",
    pages = "222--256",
    year = "2013"
}

@article{Huber:2022ndk,
    author = "Huber, S. J. and Mimasu, K. and No, J. M.",
    title = "{Baryogenesis from spontaneous CP violation in the early Universe}",
    eprint = "2208.10512",
    archivePrefix = "arXiv",
    primaryClass = "hep-ph",
    month = "8",
    year = "2022"
}

\end{document}